\documentclass[prd,showpacs,preprint,preprintnumbers,nobibnotes,superscriptaddress,floatfix]{revtex4-1}
\usepackage{amsmath,amssymb,amsbsy,color}
\usepackage{epsfig}
\usepackage{epstopdf}
\usepackage{grffile} 
\usepackage{psfrag}
\usepackage[normalem]{ulem}
\usepackage{hyperref}

\def\red{}

\def\simge{
    \mathrel{\rlap{\raise 0.511ex
        \hbox{$>$}}{\lower 0.511ex \hbox{$\sim$}}}}
\def\simle{
    \mathrel{\rlap{\raise 0.511ex 
        \hbox{$<$}}{\lower 0.511ex \hbox{$\sim$}}}}

\begin{document}

\title{Neutron and proton electric dipole moments from $N_f=2+1$ domain-wall fermion lattice QCD}

\author{Eigo Shintani}
\email{shintani@riken.jp}
\affiliation{
  RIKEN Advanced Institute for Computational Science, Kobe, Hyogo 650-0047, Japan
}
\affiliation{
  RIKEN-BNL Research Center, Brookhaven National Laboratory, Upton, NY 11973, USA
}

\author{Thomas Blum}
\affiliation{
  Physics Department, University of Connecticut, Storrs, CT 06269-3046, USA
}
\affiliation{
  RIKEN-BNL Research Center, Brookhaven National Laboratory, Upton, NY 11973, USA
}

\author{Taku Izubuchi}
\affiliation{
  RIKEN-BNL Research Center, Brookhaven National Laboratory, Upton, NY 11973, USA
}
\affiliation{
  High Energy Theory Group, Brookhaven National Laboratory, Upton, NY 11973, USA
}

\author{Amarjit Soni}
\affiliation{
  High Energy Theory Group, Brookhaven National Laboratory, Upton, NY 11973, USA
}


\collaboration{RBC and UKQCD collaborations}

\begin{abstract}
We present a lattice calculation of the neutron and proton electric dipole moments (EDM's)
with $N_f=2+1$ flavors of domain-wall fermions. 
The neutron and proton EDM form factors are extracted from three-point functions 
at the next-to-leading order in the $\theta$ vacuum of QCD. 
In this computation, we use pion masses 0.33 and 0.42 GeV and 2.7 fm$^3$ lattices with Iwasaki gauge action 
and a 0.17 GeV pion and 4.6 fm$^3$ lattice with I-DSDR gauge action, 
all generated by the RBC and UKQCD collaborations. 
The all-mode-averaging technique enables an efficient and high statistics calculation. 
Chiral behavior of lattice EDM's is discussed in the context of baryon chiral perturbation theory. 
In addition, we also show numerical evidence on relationship of three- and two-point correlation function 
with local topological distribution.
\end{abstract}
\maketitle

\section{Introduction}

Electric dipole moments (EDM) are sensitive observables of the
CP-violating (CPV) effects of the fundamental interactions described by the standard model (SM) and theories beyond the SM (BSM). 
The measurement of the neutron EDM (nEDM) has been attempted 
in experiments since the 1950's; however no evidence for the nEDM has been found, and the latest experimental upper bound is tiny,
$d_{N}\le2.9\times 10^{-26}$ e$\cdot$cm (90\% CL)\cite{Baker:2006ts,Baker:2007df}.
From the theoretical point of view, 
the contribution to the nEDM from the CPV phase in the CKM mixing matrix 
is extremely small since the first non-vanishing contribution appears at three loops, and 
$d_N\sim 10^{-31}$ e$\cdot$cm \cite{Mannel:2012hb,Khriplovich:1981ca,Khriplovich:1985jr,Czarnecki:1997bu}, 
more than 5 orders of magnitude below the experimental bound.
On the other hand, since the QCD Lagrangian contains a CP-odd $\theta$ term,  
the CPV effect from the strong interaction may dominate, even though its contribution appears to be unnaturally small, 
$d_N/\bar\theta \sim 10^{-17}$ e$\cdot$cm
\cite{Crewther:1979pi,Abada:1990bj,Aoki:1990zz,Cheng:1990pi,Abada:1991dv,Pich:1991fq,
Pospelov:1999mv,Hisano:2012sc,Borasoy:2000pq,Ottnad:2009jw,Guo:2012vf,O'Connell:2005un,
Kuckei:2005pg,Mereghetti:2010kp}. This is known as the strong CP problem.


For search of the new physics due to BSM scenarios, nEDM is just about the most important observable, 
since naturalness arguments strongly suggest that BSM interactions will not be
aligned with the usual quark mass eigenstates~\cite{Agashe:2004cp}.
As a consequence, in most BSM scenarios, there will be additional CP-odd phases, thus
nEDM is a unique way to search the effect of these new phase(s).
Extensions of the SM can generate nEDM at 1-loop order in the new interactions, 
for example Left-Right Symmetric models~\cite{Beall:1981zq},
extra-higgses models, warped models of flavor~\cite{Agashe:2004cp}
and supersymmetric (SUSY) models
~\cite{Abel:2001vy,Hisano:2004tf,Ellis:2008zy,Li:2010ax,Ibrahim:2007fb,Fukuyama:2012np}.
Indeed some of the most popular models, {\it e.g.} SUSY, have a problem
that the expected size of nEDM value is bigger than existing bounds~\cite{Engel:2013lsa}.
In fact, warped models which are considered extremely attractive for a geometric
understanding of flavors, the nEDM naturally arises around the same level as the current experimental bound,
so there is a mild tension by factors of a few.
This means that if nEDM is not discovered after another order of magnitude improvement is made,
then that will cause a serious constraint on the warped models of flavor.
To extract BSM effects arising in an EDM,
both high energy particle contributions and low energy hadronic effects have to be taken into account. 
Although there have been several estimates of BSM contributions to EDM's, 
for instance from quark electric dipole, chromoelectric dipole, and Weinberg operators, 
based on effective models, baryon chiral perturbation theory (BChPT) and sum rules
\cite{Pospelov:1999mv,Hisano:2012sc,Borasoy:2000pq,Ottnad:2009jw,Guo:2012vf,O'Connell:2005un,
Kuckei:2005pg,Mereghetti:2010kp,Bsaisou:2014oka,Bsaisou:2014zwa,Mereghetti:2015rra}, 
it is necessary to evaluate the unknown low-energy constants appearing in such models. 
On the other hand, computations from first principles using lattice QCD are also doable.
A recent attempt at estimate of quark EDM contribution
is given in \cite{Bhattacharya:2015rsa,Bhattacharya:2015esa}.

This paper presents a first step in a feasibility study of the non-perturbative 
computation of nucleon EDM's.
The starting point is to perform the path-integral 
from an {\it ab-initio} calculation including the $\theta$-term.
The renormalizability of the $\theta$-term allows a Monte-Carlo integration 
without considering the mixing with lower-dimensional CP violating operator.
It is also an appropriate test for the next step towards
inclusion of higher dimensional CP-odd sources associated with BSM theories.
Currently there are three strategies for neutron and proton EDM computations 
in lattice QCD: 

(1) Extraction of the EDM using an external electric field
   \cite{Aoki:1989rx,Aoki:1990ix,Shintani:2006xr,Shintani:2008nt,Shindler:2015aqa}, 

(2) Direct computation of the EDM form factor, in which the EDM is
   given in the limit of zero momentum transfer 
   \cite{Shintani:2005xg,Berruto:2005hg,Alexandrou:2015spa},

(3) Use of imaginary $\theta$ and extraction of the EDM as in (1) or (2).
   \cite{Izubuchi:2008mu,Aoki:2008gv,Guo:2015tla}

In (1) the neutron and proton EDM are evaluated 
from the energy difference of nucleons with spin-up and spin-down in a constant external electric field.
In~\cite{Shintani:2006xr,Shintani:2008nt} the calculation
is carried out with Minkowskean electric field, with a signal appearing as a linear response
to the magnitude of the electric field.
However, as shown in \cite{Shintani:2006xr,Shintani:2008nt}, possibly large excited state contamination 
results due to enhanced temporal boundary effects of the Minkowskean electric field. 

(2) is a straightforward method in which the EDM appears as the non-relativistic limit
of the CP violating part of the matrix element of the 
the electromagnetic (EM) current in the ground state of the nucleon.
It requires the subtraction of CP-odd contributions arising from mixing of the CP-even and
odd nucleon states in the $\theta$-vacuum~\cite{Shintani:2005xg,Berruto:2005hg}.
In this method, the EDM is obtained from the form factor at zero momentum transfer. 
This paper employs this strategy. 

In (1) and (2), the $\theta$-term in Euclidean space-time is pure imaginary while the CP-even part
of the action is real, which leads to a so-called sign problem for Monte-Carlo simulation.
To avoid this issue, the idea of (3) is to employ a purely real action 
by using an imaginary value of $\theta$ in the generation of gauge field configurations. 
This has an advantage of improved signal-to-noise over the reweighting method. 
In~\cite{Izubuchi:2008mu,Aoki:2008gv} preliminary results indicate 
relatively small statistical errors for the nEDM, however we note that
these results may be affected by lattice artifacts due chiral symmetry breaking of Wilson-type fermions.
Recently updated results in $N_f=2+1$ QCD using (3) have been presented in Ref.~\cite{Guo:2015tla}
and appear promising.

Figure \ref{fig:dn} (also see \cite{Shintani:ConfinementX2012}) shows the summary plot
of EDM results obtained using the strategies (1) and (3) and Wilson-clover fermions and strategy (2) using
domain-wall fermions (DWF) which maintain chiral symmetry at non-zero lattice spacing to a high degree
\cite{Furman:1994ky}.
Older results suffer from large statistical errors and uncontrolled systematic errors. 
To pursue a more reliable estimate of the neutron and proton EDM's, 
we adopt strategy (2) and use DWF. To efficiently reduce statistical errors
we employ all-mode-averaging (AMA)~\cite{Blum:2012uh,Blum:2012my,Shintani:2014vja}.

This paper is organized as follows:
in section \ref{sec:mes_EDM} we introduce notation and give formulae used to extract 
the CP-even EM and CP-odd EDM form factors for the neutron and proton from correlation
functions computed in lattice QCD. 
In section \ref{sec:results} we first describe the lattice setup, including AMA 
parameters, and then give numerical results for the EM and EDM form factors and 
subsequent neutron and proton EDM's. 
We discuss our lattice QCD result in the context of phenomenological estimates 
in section \ref{sec:discuss} and present an idea to further reduce statistical errors related to reweighting in section~\ref{sec:idea}. Finally we summarize our study in~\ref{sec:summary}. 

\section{Measurement of EDM form factor}
\label{sec:mes_EDM}

\subsection{Extraction of EDM form factor}

The matrix element is parameterized similarly, with CP-even and odd form factors,
\begin{eqnarray}
&&\langle N(\vec p_f,s_f)|V_\mu^{\rm EM}|N(\vec p_i,s_i)\rangle_\theta
  = \bar u_N^\theta(\vec p_f,s_f)\Big[ F_1(q^2)\gamma_\mu + \frac{i F_2(q^2)}{2 m_{N}} \frac{[\gamma_\mu,\gamma_\nu]}{2} q_\nu
  \nonumber\\
&&+ \frac{F_3(q^2)}{2 m_{N}}\frac{\gamma_5[\gamma_\mu,\gamma_\nu]}{2}q_\nu
  \Big]u_N^\theta(\vec p_i,s_i).
\label{eq:matrix_elem}
\end{eqnarray}
where $F_{1}$ and $F_2$ are the usual CP-even EM form factors, 
and $F_3$ is the CP-odd EDM form factor. 
Here we focus on the electromagnetic interaction with quarks inside nucleon under $\theta$-vacuum, 
and so that $\langle\rangle_\theta$ is explicit representation of path-integral 
with $\theta$-term.
$u_N^\theta$ denotes the nucleon spinor-function as a function of $\theta$. 
Each form factor is able to be extracted from 
order-by-order in $\theta$ in the expanded three-point function and 
Eq.~(\ref{eq:matrix_elem}) as shown below
(also see \cite{Shintani:2005xg,Berruto:2005hg} for more detail). 
Note that momentum transfer $q=p_f-p_i$ is used in the space-like region.

We represent the three-point function in our lattice study as
\begin{eqnarray}
  C_{V_\mu}^\theta(t_f,\vec p_f;t, \vec q;t_i, \vec p_i) &=& C_{V_\mu}(t_f,\vec p_f;t, \vec q;t_i, \vec p_i) 
  \nonumber\\
&+& i\theta C_{V_\mu}^{Q}(t_f,\vec p_f;t, \vec q;t_i, \vec p_i) + O(\theta^2),
\label{eq:3pt_theta}
\end{eqnarray}
where all terms on the RHS are computed in the $\theta=0$ vacuum, but the second is
reweighted with topological charge $Q=\int G \tilde G/64\pi^2$ using gluon field strength $G$,
in QCD action with $\theta$ term, $S_{\rm QCD}+i\theta Q$.
Here the EM current is defined by the local bilinear,
$V_\mu^{\rm EM}=Z_V\bar q\gamma_\mu Q_cq$ 
with quark charge matrix $Q_{c}={\rm diag}(2/3,-1/3,-1/3)$, as in the continuum theory, but
multiplied by the lattice renormalization factor $Z_V$.
In this paper, we ignore the SU$_f$(3) suppressed disconnected quark diagrams
and compute only the connected part in three-point function.

We use the following ratio,
\begin{eqnarray}
  R_\mu(t_f,\vec p_f;t, \vec q;t_i, \vec p_i) = K
  \frac{C_{V_\mu}(t_f,\vec p_f;t, \vec q;t_i, \vec p_i)}{C_G(t_f-t_i,\vec p_f)}\bigg[ 
  \frac{C_L(t_f-t,\vec p_i)C_G(t-t_i,\vec p_f)C_L(t_f-t_i,\vec p_f)}
       {C_L(t_f-t,\vec p_f)C_G(t-t_i,\vec p_i)C_L(t_f-t_i,\vec p_i)}\bigg]^{1/2}
\end{eqnarray}
where $K=\sqrt{(E_N(\vec p_f)+m_N)(E_N(\vec p_i)+m_N)}/\sqrt{E_N(\vec p_f)E_N(\vec p_i)}$. 
The nucleon two-point function
with smeared-source/smeared-sink is $C_G(t,\vec p)$ 
and smeared-source/local-sink is $C_L(t,\vec p)$. 
Taking the large time-separation limit to project onto the nucleon ground states,
\begin{eqnarray}
&&\mathcal R_\mu(t_f,\vec p_f;t, \vec q;t_i, \vec p_i) \equiv 
  \lim_{t_f-t,t-t_i\rightarrow\infty}R_\mu(t_f,\vec p_f;t, \vec q;t_i, \vec p_i) 
  \nonumber\\
&&= \sum_{s_f,s_i}u_N^\theta(\vec p_f,s_f)
  \langle N(\vec p_f,s_f)|V_\mu|N(\vec p_i,s_i)\rangle_\theta \bar u_N^\theta(\vec p_i,s_i)
  \nonumber\\
&&= \mathcal R_\mu(\vec p_f,\vec p_i) + i\theta\mathcal R^Q_\mu(\vec p_f,\vec p_i) 
    + \mathcal O(\theta^2),
\label{eq:ratio_lim}
\end{eqnarray}
for the matrix element in (\ref{eq:matrix_elem}).

To describe the RHS of (\ref{eq:ratio_lim})
up to the second order in $\theta$, we replace the spinor sums  by
the matrix \cite{Shintani:2005xg}
\begin{eqnarray}
  \sum_s u_N^\theta(\vec p,s)\bar u^\theta_N(\vec p,s) 
   &=& E_N\gamma_0 - i\vec p\cdot\vec\gamma + m_Ne^{i\alpha_N(\theta)\gamma_5},\\
   &\approx& E_N\gamma_0 - i\vec p\cdot\vec\gamma + m_N (1+i\alpha_N(\theta)\gamma_5)
   + \mathcal O(\theta^2),
\label{eq:spin sum}   
\end{eqnarray}
where the CP-odd mixing angle $\alpha_N(\theta)$ induced by the $\theta$-term appears explicitly
Here $\alpha_N(\theta)$ is a Lorentz scalar, thus it is as a function of quark mass. 
To the lowest order, $\alpha_N(\theta)\approx \theta\alpha_{N}$ 
is determined by 
\begin{equation}
  {\rm tr}\Big[ \gamma_5C_{L/G}^\theta(t,\vec p)\Big] 
  \simeq {Z^*}_{L/G}Z_G\frac{2m_N}{E_N}i\alpha_N\theta\big(e^{-E_Nt}+(-)^be^{-E_N(L_t-t)}\big),
  \label{eq:alpha_N}
\end{equation}
in enough large $t$. 
$Z_{L/G}$ denotes normalization factor for local (L) or Gaussian smeared (G) sinks.
$b$ indicates the boundary condition in the temporal direction
with size $L_t$; $b=0$ is for periodic boundary conditions, and 
$b=1$ anti-periodic. $N^{*}$ denotes the parity partner of the nucleon in the $\theta=0$ vacuum. 
Note that to the order we are working, $Z$'s and $E$'s are 
given by the usual lowest order of $\theta$, CP-even quantities.

Using (\ref{eq:spin sum}) and the definitions in (\ref{eq:matrix_elem}),
and taking traces with projectors $P_{4}^{+}\equiv(1+\gamma_{4})/2$ and
$P_{5z}^{+}\equiv i(1+\gamma_{4})\gamma_{5}\gamma_{z}/2$, 
the leading order in $\theta$ ($\theta$-LO) form factors are obtained from (\ref{eq:ratio_lim}) by
\begin{eqnarray}
{\rm tr}\Big[ P_{5z}^+\mathcal R_{x}(0,\vec p) \Big] 
&=& \frac{p_{y}}{E_N}G_m(q^2), 
\label{eq:gm}\\
{\rm tr}\Big[ P_{5z}^+\mathcal R_{y}(0,\vec p) \Big] 
&=& -\frac{p_{x}}{E_N}G_m(q^2), 
\label{eq:gmy}\\
{\rm tr}\Big[ P_4^+\mathcal R_t(0,\vec p) \Big] 
&=& \frac{E_N+m_N}{E_N}G_e(q^2),
\label{eq:ge}
\end{eqnarray}
with Sachs electric and magnetic form factors
\begin{equation}
  G_e(q^2) = F_1(q^2) - \frac{q^2}{4m_N} F_2(q^2),\quad
  G_m(q^2) = F_1(q^2) + F_2(q^2).
\end{equation}

Similarly, including the $\alpha_{N}$ term in (\ref{eq:spin sum}),
the form factors appearing at next-to-leading order in $\theta$ ($\theta$-NLO) are obtained from
\begin{eqnarray}
{\rm tr}\Big[ P_{5z}^+\mathcal R^Q_{t}(\vec p_f,\vec p_i)\Big]
&=& i\frac{p_z}{2E_N}\bigg[\alpha_N\bigg\{ F_1(q^2) + \frac{3m_N+E_N}{2m_N}F_2(q^2)\bigg\}
- \frac{E_N+m_N}{m_N}F_3(q^2)\bigg].
\label{eq:rQ_t}
\end{eqnarray}
The EDM form factors $F_3$ are then determined by  the subtracting the $\alpha_N F_{1,2}$ terms.

\section{Numerical results}
\label{sec:results}
\subsection{Lattice parameters}

We use lattices with size $L_\sigma\times L_t = 24^3\times 64$, Iwasaki gauge action 
with $a^{-1}=1.7848(6)$ GeV (gauge coupling is $\beta=2.13$) \cite{Aoki:2010dy},
and $L_\sigma\times L_t = 32^3\times 64$, 
Iwasaki(I)-DSDR gauge action with $a^{-1}=1.3784(68)$ GeV (gauge coupling is $\beta=1.75$) 
\cite{Arthur:2012opa}. 
Both lattice scales were determined from a global, continuum and chiral fit~\cite{Blum:2014tka}, 
including physical point ensembles.
The fermions are domain wall fermions (DWF), which significantly suppresses the $\mathcal O(a)$
lattice artifact due to chiral symmetry breaking. 
The additive quark mass shift from the explicit chiral symmetry breaking, or residual mass, is 
$am_{\rm res}=0.0032$ and $am_{\rm res}=0.0019$ 
for the Iwasaki $24^3$ and I-DSDR $32^3$ ensembles, respectively. 
The chiral symmetry of domain-wall fermions is useful to investigate 
the chiral behavior of the EDM without any additive renormalization. 
We use the two light quark masses $m=0.005$ and $m=0.01$, corresponding to 
330 and 420 MeV pion mass for the Iwasaki 24$^3$ ensembles, and 
$m=0.001$ corresponding to a 170 MeV pion mass for I-DSDR 32$^3$ ensemble, 
in order to investigate the chiral behavior of nucleon EDM. 
To suppress correlations between measurements on successive configurations, 
we use a 10 (unit length) trajectory separation for Iwasaki 24$^3$ and 
16 trajectory separation for I-DSDR 32$^3$. 
The renormalization factor for the vector current is given as
$Z_V=0.71273(26)$ for Iwasaki 24$^3$ \cite{Blum:2014tka}, 
and $Z_V=0.6728(80)$ for I-DSDR 32$^3$ \cite{Arthur:2012opa}.
Both are evaluated at $-m_{\rm res}$, $i.e.$, in the chiral limit.
Table~\ref{tab:param} shows the lattice parameters on each gauge ensemble. 

\begin{table}
\begin{center}
\caption{Lattice and AMA parameters. $N_G$ refers to
  the number of AMA measurements per configuration and $N_\lambda$ the number of eigenvectors.
}\label{tab:param}
\begin{tabular}{ccccccccccc}
\hline\hline
Size & $a^{-1}$(GeV) &Vol.(fm$^3$)& $L_s$ & mass & configs & $N_G$ & $N_\lambda$ 
& AMA approx & $m_\pi$(MeV) & $t_{\rm sep}$(fm)\\
\hline
$24^3\times 64$ &1.7848(6) &2.7$^3$& 16 & 0.005 & 32 & 400 & $|r|<0.003$ & 330 & 772 & 1.32\\
& & & & & & & & & 187 & 0.9 \\
$24^3\times 64$ &1.7848(6) &2.7$^3$& 16 & 0.01  & 32 & 180 & $|r|<0.003$ & 420 & 701 & 1.32\\
& & & & & & & & & 133 & 0.9 \\
$32^3\times 64$ &1.3784(68) &4.6$^3$& 32 & 0.001 & 39 & 112 & 1000 & 100-125 CG iter & 170 & 1.29 \\
\hline\hline
\end{tabular}
\end{center}
\end{table}

We use Gaussian-smeared sources as described in \cite{Yamazaki:2009zq} 
with width 0.7 for Iwasaki $24^{3}$ and 0.6 for I-DSDR $32^{3}$ ensembles, respectively, 
and the number of hits of the 3D Laplacian was 100 and 70, respectively.
The three-point function is constructed with a zero-spatial-momentum sequential source ($\vec p_f=0$)
on a fixed time-slice for the sink nucleon operator (see~\cite{Sasaki:2003jh} for details).
Fourier transforming the position of the EM current  injects
spatial momentum $\vec q=\vec p$, so $\vec p_i=-\vec p$ is removed at the source 
by momentum conservation. 
In this analysis we employ four different spatial momentum-transfer-squared values, 
$|\vec q|^2=4\pi^2\vec n_p^2/L^2_\sigma,\, \vec n_p^2=1,2,3,4$,
and average over all equivalent values of $|\vec p|^2$ to improve statistics.
The Euclidean time-separation of the sink and source in the three-point function 
is set to 12 and 9 time-slices for $24^{3}$ and $32^{3}$ ensembles, respectively
(both about 1.3 fm).
On Iwasaki $24^{3}$ we also employ a shorter separation of 8 time slices 
to investigate excited state contamination.

The AMA parameters \cite{Blum:2012uh,Blum:2012my,Shintani:2014vja}
we used here are also in Table~\ref{tab:param}. 
Here translational invariance is employed as the covariant symmetry to be averaged over.
Approximate quark propagators on each time slice are computed  
starting from the initial source locations and shifting once in each direction 
by one-half of the spatial linear size of the lattice. 
In addition, on I-DSDR $32^{3}$ ensemble, we repeat three more times, 
starting from a different initial spatial source location
(except for 16 source locations).
To compute the bias correction, 
the exact (to numerical precision) propagators are computed at 
the same initial source location(s) on one time-slice for 24$^3$ or 
each time-slice for 32$^3$. 

Quark propagators are computed using the conjugate gradient (CG) algorithm and 
the 4D-even-odd-preconditioned Dirac operator~\cite{Blum:2012uh,Blum:2012my,Shintani:2014vja}. 
As shown in Table~\ref{tab:param}, 
we compute the various lowest modes of the preconditioned operator to deflate 
the CG and to construct the approximate quark propagators using  
the implicitly restarted Lanczos algorithm with Chebyshev polynomial 
acceleration \cite{Neff:2001zr}.
Especially, for I-DSDR $32^3$ ensemble, a M\"obius Dirac operator with $L_s=16$ was used 
for the approximation instead of the DWF operator with $L_s=32$ 
to reduce the memory footprint
\cite{Brower:2012vk,Yin:2011np,Blum:2014wsa}.
In addition, the eigenvectors for this case were computed in mixed precision 
and stored in single precision.
In Reference \cite{Shintani:2014vja} a detailed discussion of these AMA procedures and 
the attendant bias is discussed.

\subsection{Topological charge distribution}

\begin{figure}[t]
\begin{center}
  \includegraphics[width=100mm]{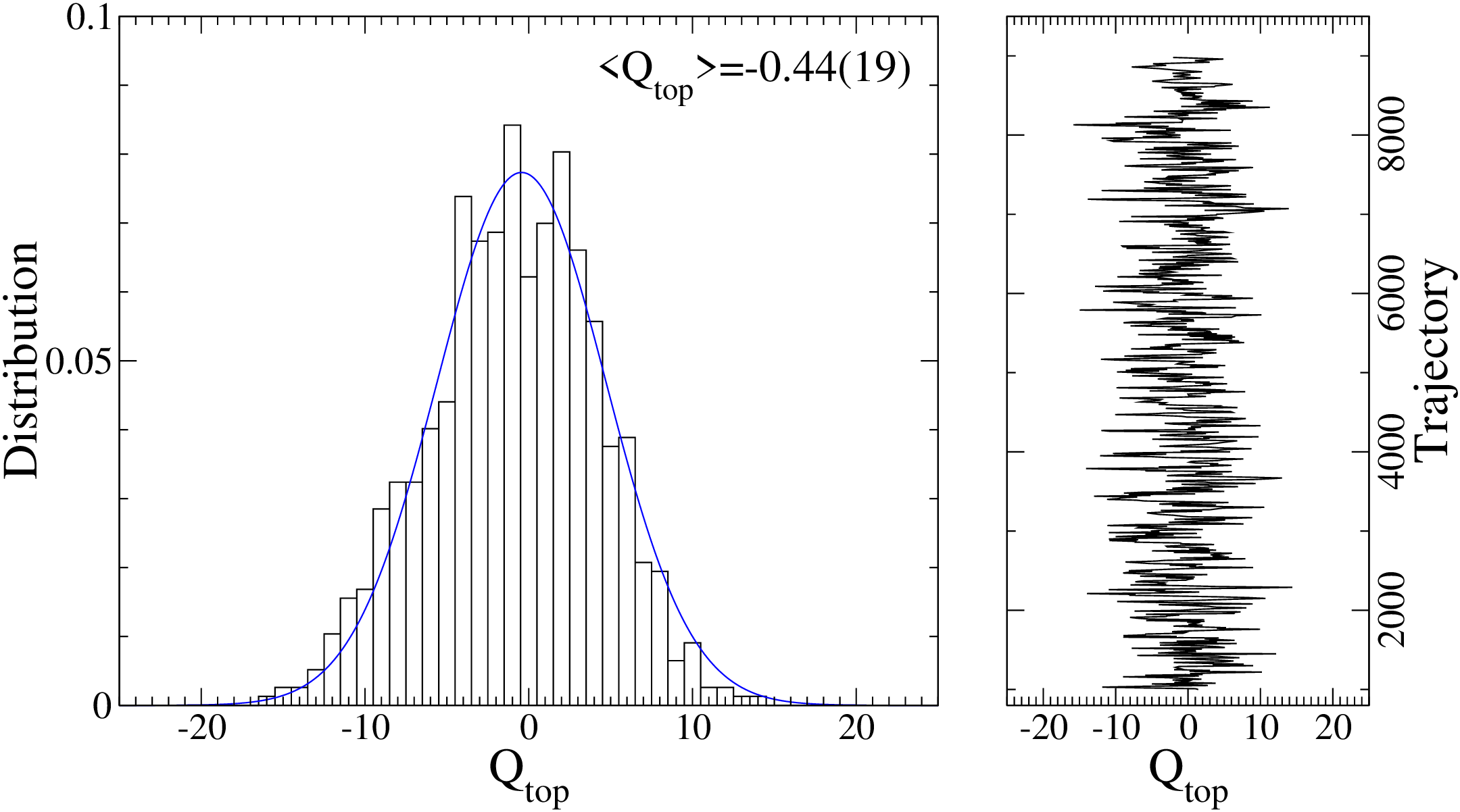}
  \vskip 3mm
  \includegraphics[width=100mm]{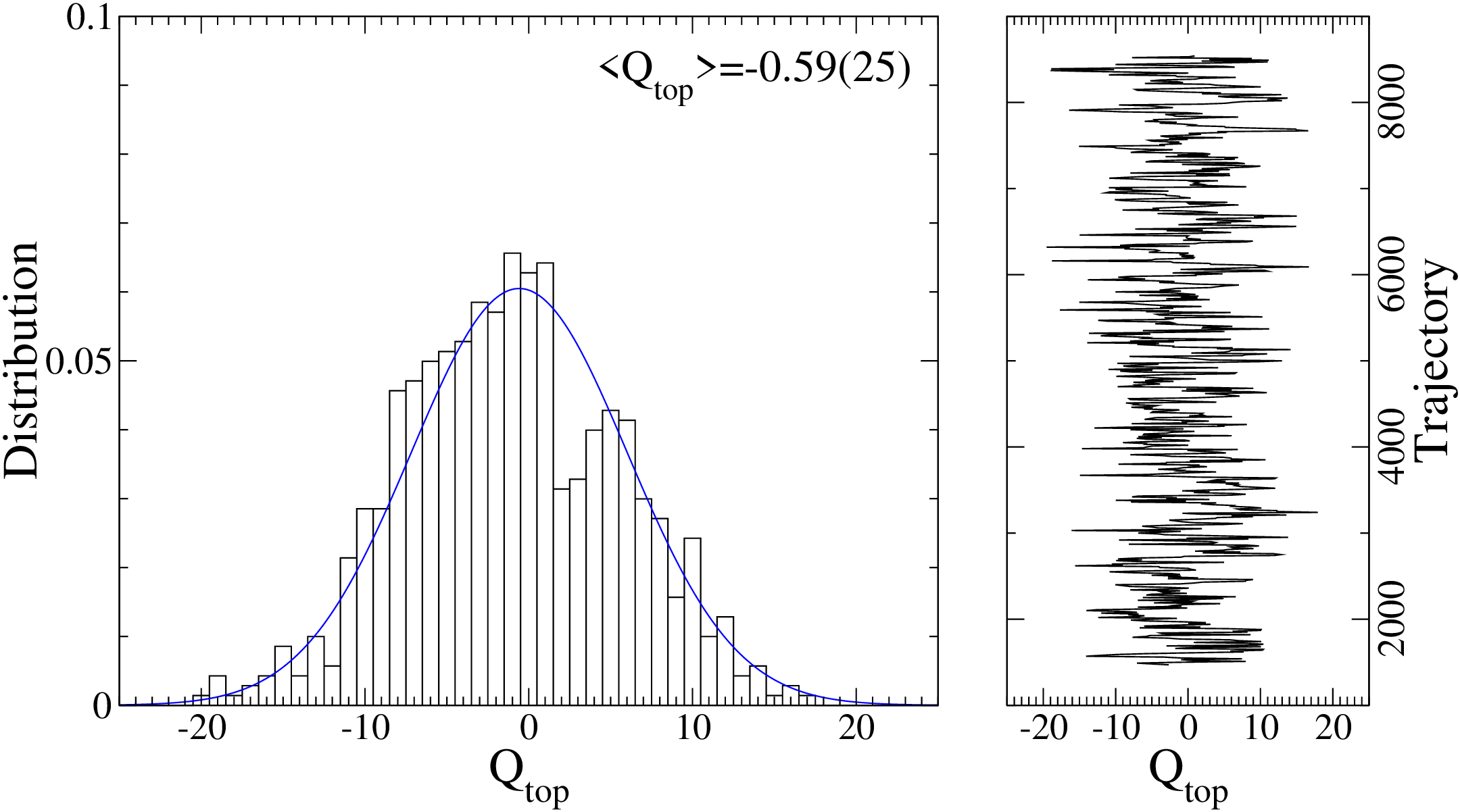}
  \caption{Distribution of topological charge and its Monte Carlo time history.
  Pion mass 330 MeV (top) and 420 MeV (bottom), Iwasaki 24$^3$, ensembles. 
  The solid line represents a Gaussian distribution function.
  }
  \label{fig:Qdist}
\end{center}
\end{figure}

\begin{figure}[t]
\begin{center}
  \includegraphics[width=100mm]{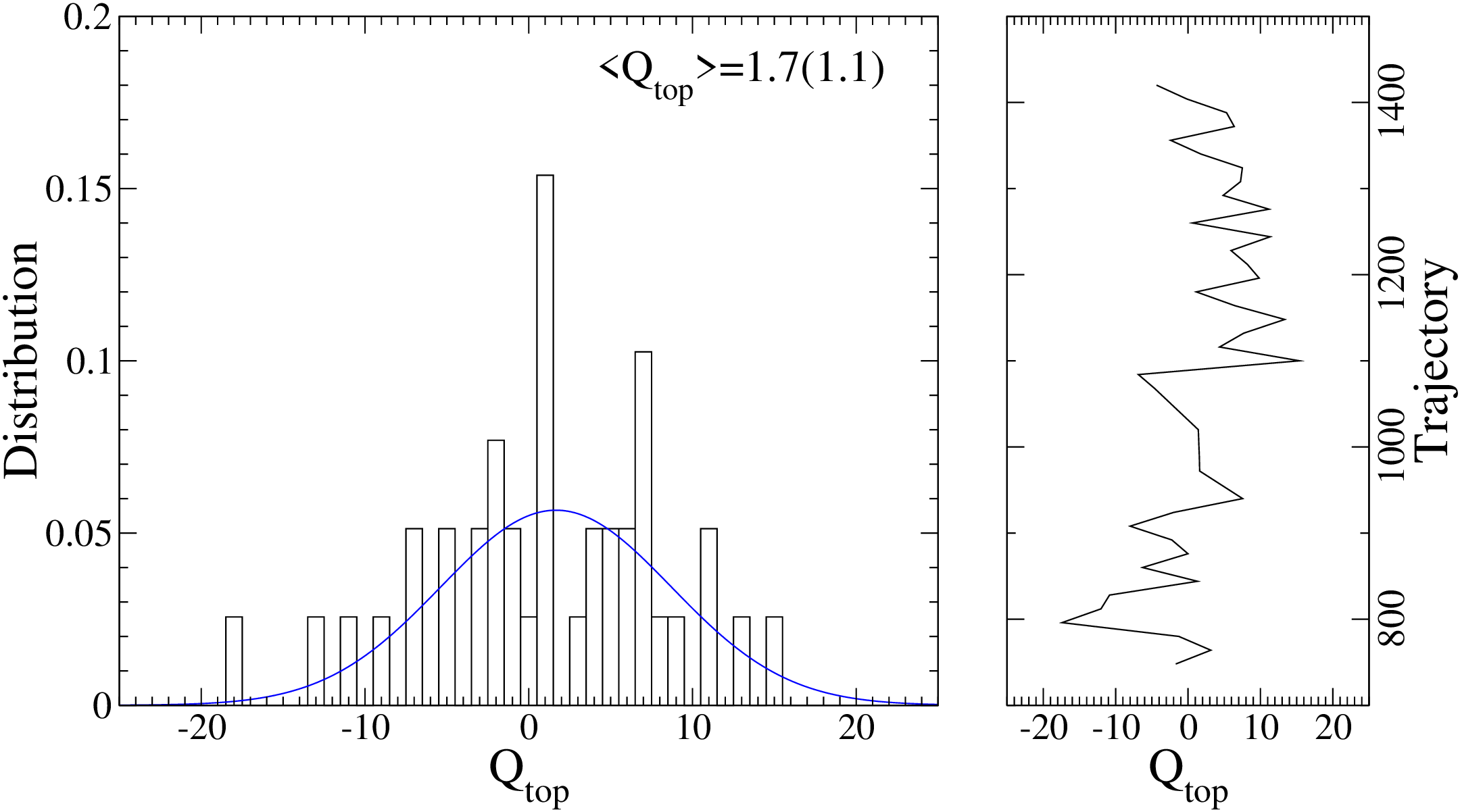}
  \caption{Same as Figure \ref{fig:Qdist} but for the I-DSDR $32^3$ ensemble in 170 MeV pion.}
  \label{fig:QdistID}
\end{center}
\end{figure}

We describe the topological charge distribution used in our analysis 
of the CP-odd parts of the two- and three-point functions. 
Topological charge $Q$ is computed using the 5-loop-improved 
lattice topological charge~\cite{de_Forcrand:1997sq}
which is free of lattice spacing discretization errors through $\mathcal O(a^4)$.
The gauge fields are smoothed before computing $Q$ by 
APE smearing~\cite{Falcioni:1984ei,Albanese:1987ds} 
with smearing parameter 0.45 for 60 sweeps.
Figures \ref{fig:Qdist} and \ref{fig:QdistID} show histograms of the topological charge 
and its Monte Carlo time history for the ensembles used here. 
The shape is roughly Gaussian for the Iwasaki 24$^3$ ensembles, on the other hand 
for the I-DSDR 32$^3$ there is significant deviation from zero 
where measurements were made on only 39 configurations 
(the distribution for the whole ensemble looks much better~\cite{Arthur:2012opa}). 
Despite the poor shape, at least the peak is near $Q=0$, and it is roughly symmetric.  
We also observe a rather long auto-correlation time of the topological charge for this ensemble. 

The topological susceptibility obtained on these ensembles is
\begin{eqnarray}
\chi_Q=\langle Q^2\rangle/V &=& \left\{\begin{array}{cl}
3.1(2)\times 10^{-4}\textrm{ GeV}^4 & (\textrm{330 MeV pion, Iwasaki 24$^3$)},\\
4.4(2)\times 10^{-4}\textrm{ GeV}^4 & (\textrm{420 MeV pion, Iwasaki 24$^3$)}, \\
0.9(2)\times 10^{-4}\textrm{ GeV}^4 & (\textrm{170 MeV pion, I-DSDR 32$^3$)},\\
\end{array}\right.
\end{eqnarray}
and one sees the suppression with quark mass expected from chiral perturbation theory 
\cite{Leutwyler:1992yt}.
$\chi_Q$ can be used to investigate 
the relationship between the axial anomaly in QCD and CP-odd effects at $\theta$-NLO 
\cite{Leutwyler:1992yt,Aoki:2007ka}, for instance the mixing angle $\alpha_N$ or the nucleon EDM.
We discuss this point later.

\subsection{Nucleon two-point function}

The values of the nucleon mass (energy) and mixing angle
$\alpha_N$ are obtained by fitting with nucleon two-point function
using a single exponential function (see Tab.~\ref{tab:Nmass}). 
The nucleon energy and wave function renormalization $Z_{L/G}$ are obtained 
from the CP-even part of the nucleon propagator ($\theta$-LO) 
using the spin-projector $P^+_4$.
$\alpha_N$ is obtained from the CP-odd part using Eq.(\ref{eq:alpha_N}).
Since we are only working to $\theta$-NLO, to reduce the statistical error on $\alpha_N$,
the mass in the CP-odd part is fixed to the $\theta$-LO mass obtained from the CP-even part.
The fit ranges are given Tab.~\ref{tab:Nmass}, and were chosen to produce a $\chi^{2}/$d.o.f
roughly equal to 1, but with as small errors as possible. 

\begin{figure}[tb]
\begin{center}
  \includegraphics[width=140mm]{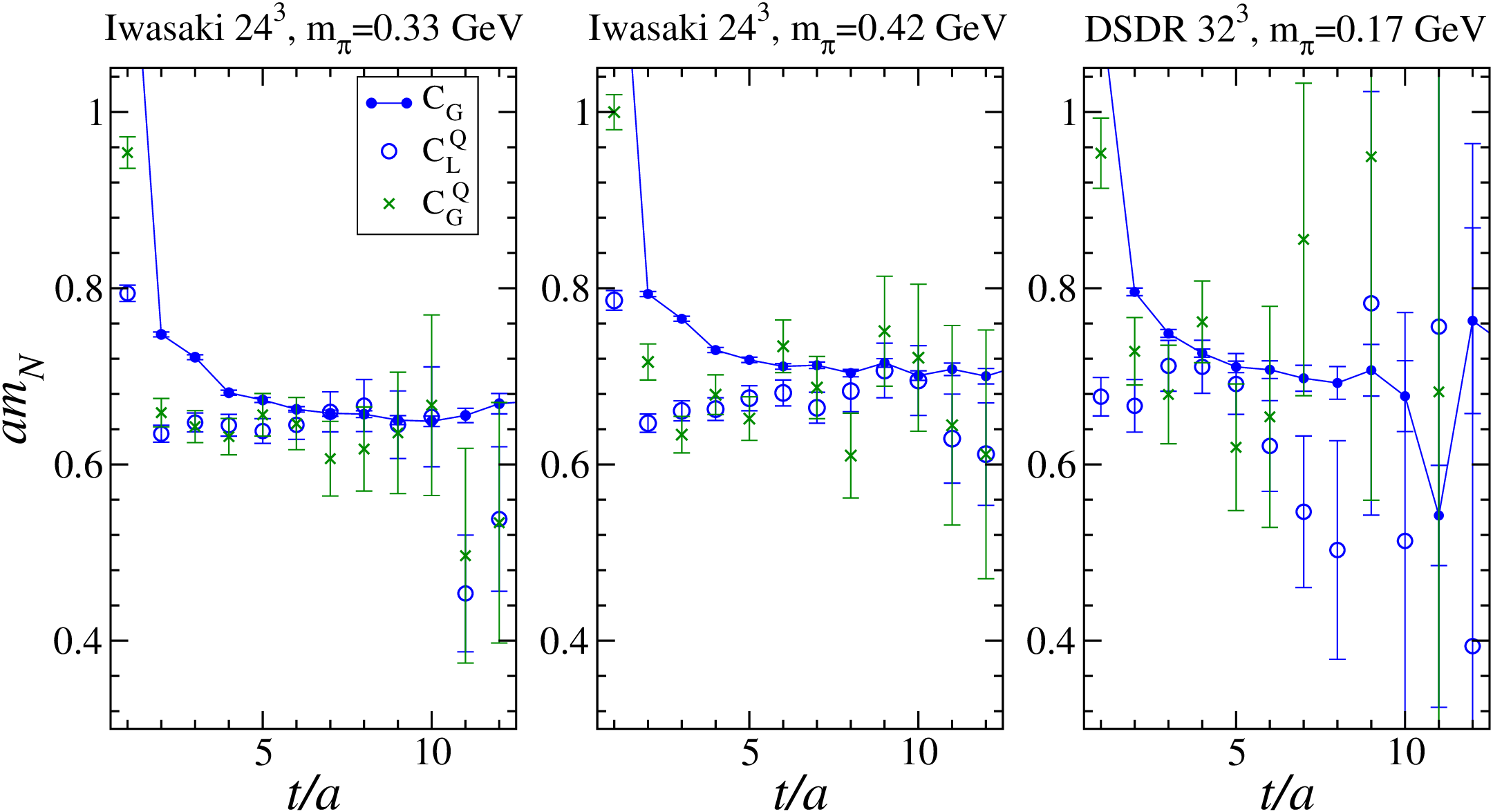}
  \caption{Effective mass of the nucleon ($\theta$-LO, Gaussian smeared sink) 
  compared to the $\theta$-NLO effective mass using local and Gaussian sinks. 
  $m_{\pi}=330$ MeV (left) and 420 MeV (middle), Iwasaki 24$^3$, and 
  170 MeV, I-DSDR 32$^3$ (right).}
  \label{fig:effm_NNG5Q}
\end{center}
\end{figure}

As shown in Fig.~\ref{fig:effm_NNG5Q}, the effective mass 
of the $\theta$-NLO nucleon propagator has a clear plateau, and its value
is consistent with that from the $\theta$-LO nucleon propagator
for both local and smeared sinks.
Plateau of effective mass plot for $\theta$-NLO seems to start at shorter time separation
than those for $\theta$-LO.
We also note the constancy of $\alpha_N$ even when the nucleon carries 
finite momentum which is in agreement with the formulation in Eq.(\ref{eq:alpha_N}). 
In the following analysis we use $\alpha_N$ computed with the Gaussian sink, 
evaluated at zero momentum.

\begin{table}[t]
\begin{center}
\caption{The nucleon energy and its CP-odd mixing angle $\alpha_N$.
The nucleon energy and $\alpha_N$ are given for the Gaussian smeared sink operator. 
}
\label{tab:Nmass}
\begin{tabular}{ccccccc}
\hline\hline
Iwasaki 24$^3$ in 0.33 GeV pion\\
\hline
fit-range & $[  6,  12]$ & $[  5,  9]$ \\
\hline
$\vec p^2$(GeV$^2$)&$E_{N}$(GeV)& $\alpha_{N}$ \\
0.000&  1.1738(25)&  -0.356(22)\\
0.218&  1.2618(27)&  -0.350(22)\\
0.437&  1.3480(34)&  -0.348(22)\\
0.655&  1.4321(52)&  -0.342(24)\\
0.873&  1.5092(90)&  -0.334(27)\\
\hline\hline
Iwasaki 24$^3$ in 0.42 GeV pion\\
\hline
fit-range & $[  7,  13]$ & $[  5,   9]$ \\
\hline
$\vec p^2$(GeV$^2$)&$E_N$(GeV)& $\alpha_{N}$ \\
0.000&    1.2641(28)&  -0.370(22)\\
0.218&    1.3454(31)&  -0.367(23)\\
0.437&    1.4210(40)&  -0.366(23)\\
0.655&    1.4931(57)&  -0.363(24)\\
0.873&    1.5660(93)&  -0.357(27)\\
\hline\hline
I-DSDR 32$^3$ in 0.17 GeV pion\\
\hline
fit-range & $[  5,  10]$ & $[  5,  9]$ \\
\hline
$\vec p^2$(GeV$^2$)&$E_{N}$(GeV)& $\alpha_{N}$ \\
0.000&  0.9746(66)&  -0.333(128)\\
0.073&  1.0122(69)&  -0.269(132)\\
0.147&  1.0491(78)&  -0.409(230)\\
0.220&  1.0827(86)&  -0.448(287)\\
0.293&  1.1116(114)&  -0.381(148)\\
\hline\hline
\end{tabular}
\end{center}
\end{table}

\subsection{Electromagnetic form factor}

First we present the CP-even form factors $G_{e}$ and $G_{m}$ obtained from
Eq.(\ref{eq:ge}) and Eqs.(\ref{eq:gm}),(\ref{eq:gmy}). 
For the Iwasaki $24^{3}$ ensembles, precise results for the (iso-vector) form factors, 
using multiple sources method, have appeared previously~\cite{Yamazaki:2009zq}.
Using AMA, we achieve a further reduction of the statistical errors compared to previous work.
The precise measurement of the EM form factors is important for the EDM calculation 
since linear combinations of $G_{e}$ and $G_m$
are needed for the subtraction terms proportional to $\alpha_{N}$. 

\begin{figure}[tb]
\begin{center}
  \includegraphics[width=125mm]{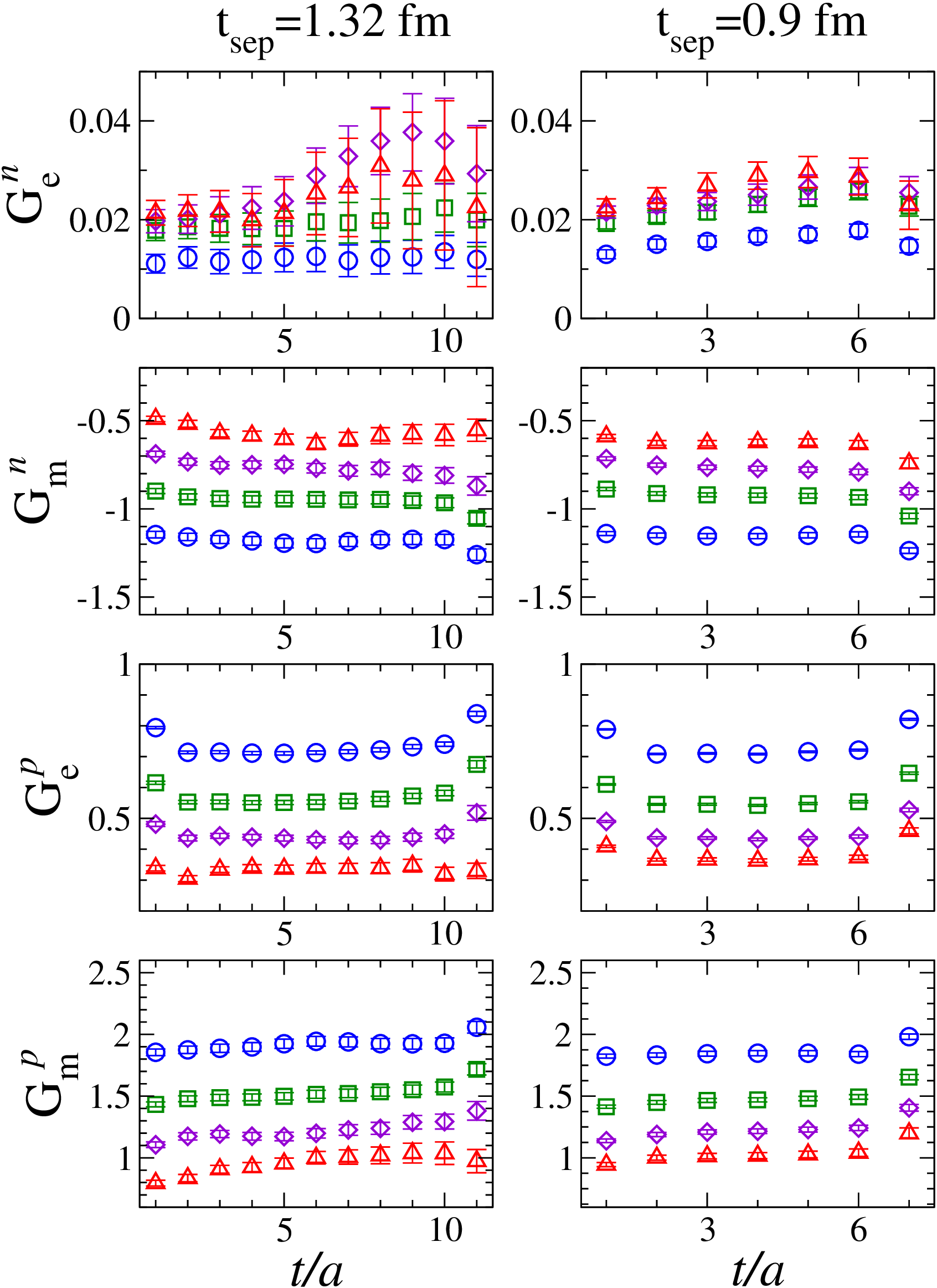}
  \caption{The operator time-slice dependence of electric and magnetic Sachs form factors 
    for the proton and neutron with $t_{\rm sep}=1.32$ fm (left) and $t_{\rm sep}=0.9$ fm (right) in
    Iwasaki 24$^3$, 330 MeV pion ensemble.
    Source and sink operators are located in $t/a=0$ and 12 ($t_{\rm sep}=1.32$ fm),
    and $t/a=0$ and 8 ($t_{\rm sep}=0.9$ fm).
    Circle, square, diamond and upper-triangle are results at $n_{\vec p}^2=1$, 2, 3, 4.
  }
  \label{fig:gem}
\end{center}
\end{figure}

\begin{figure}[tb]
\begin{center}
  \includegraphics[width=70mm]{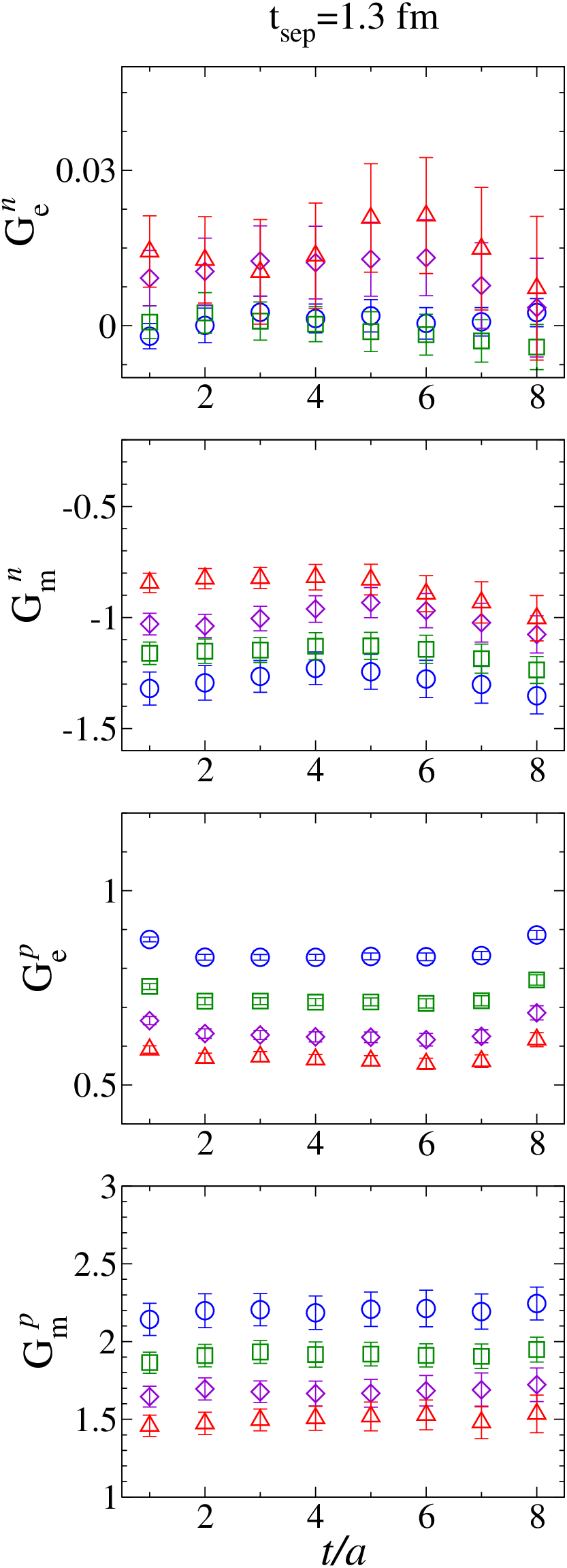}
  \caption{Same as Figure \ref{fig:gem} but for I-DSDR 32$^3$, 170 MeV pion ensemble.
  Source and sink operators are located in $t/a=0$ and 10.}
  \label{fig:gemID}
\end{center}
\end{figure}

In Figs.~\ref{fig:gem} and \ref{fig:gemID} we show the time-slice dependence 
of the EM form factors for each momenta and also compare the results for two different 
time-separations, $t_{\rm sep}$, between the nucleon source and sink operators.
Suitable nucleon ground state form factors can be extracted from the plateau regions 
$4\le t/a\le 8$, as seen in Fig.~\ref{fig:gem} (left panel) and 
$3\le t/a\le 6$ in Fig.~\ref{fig:gemID} for the smaller quark mass I-DSDR ensemble 
(note the electric form factor for the neutron is very small, and should be zero at $q^{2}=0$).
In these regions excited state contributions are evidently suppressed.
Although increasing $t_{\rm sep}$ reduces excited state contamination, 
the signal-to-noise ratio also decreases exponentially.

\begin{figure}[tb]
\begin{center}
  \includegraphics[width=110mm]{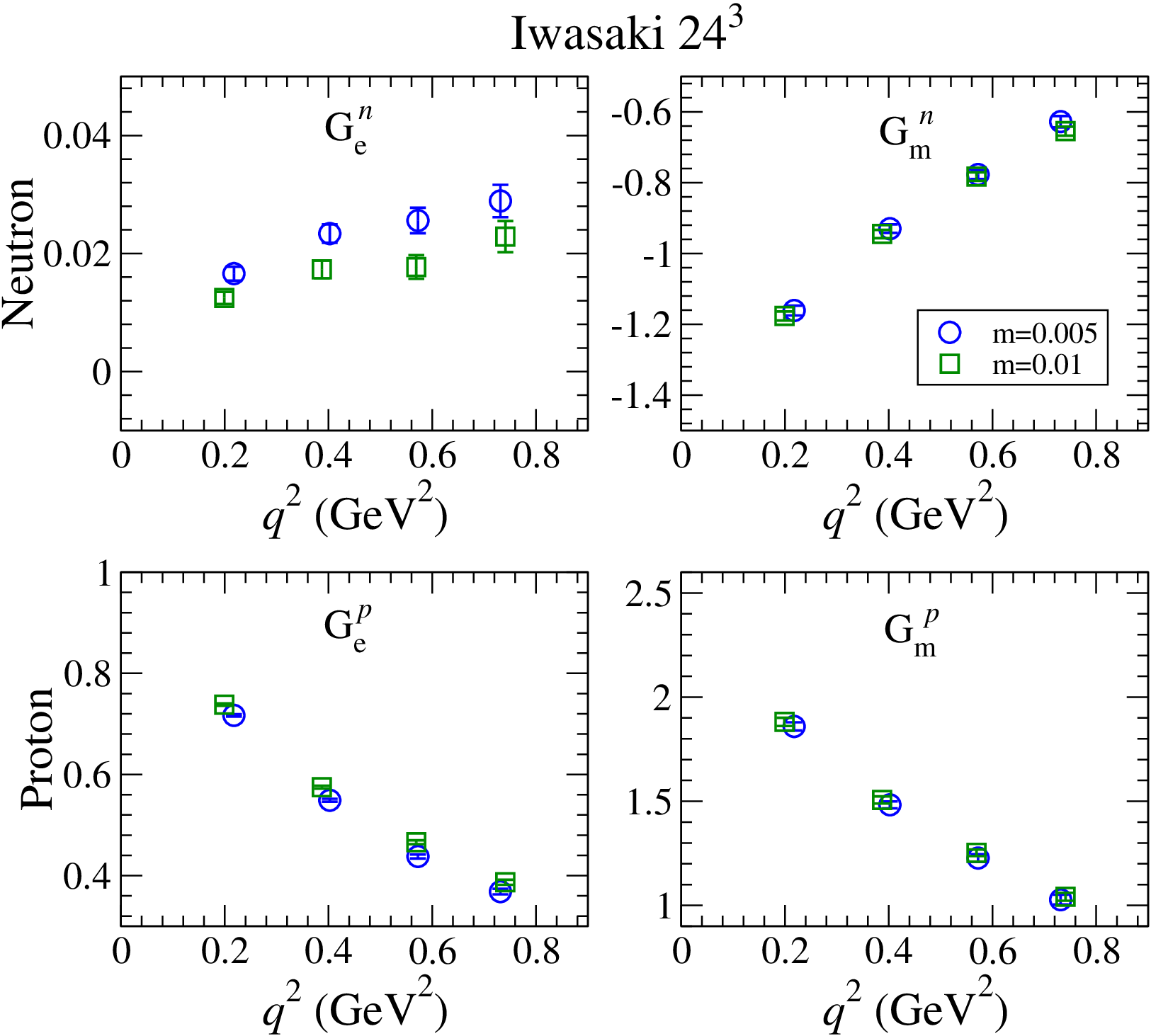}
  \vskip 5mm
  \includegraphics[width=110mm]{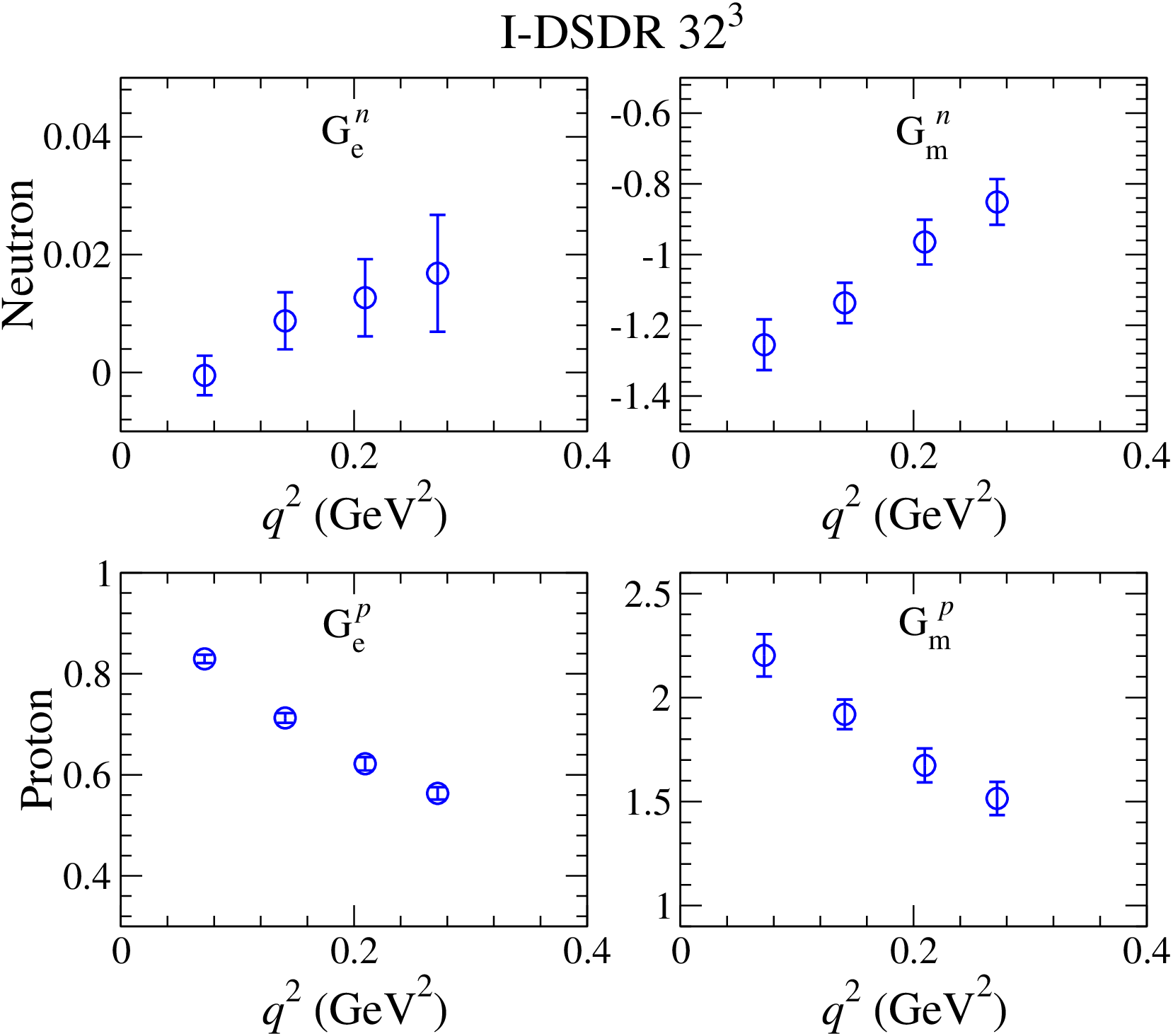}
  \caption{Electric and magnetic form factors.
    (Top) $m_\pi=330$ MeV (circle) and 420 MeV (square), 
    $t_{\rm sep}=0.9$ fm, Iwasaki 24$^3$ ensembles.
    (Bottom) I-DSDR 32$^3$, 170 MeV pion ensemble.    
  }
  \label{fig:gem_pdep}
\end{center}
\end{figure}

To see whether our value of $t_{\rm sep}$ is large enough,
we compare the form factors computed using two different values on the $24^{3}$ ensembles.
In the right panel of Fig.~\ref{fig:gem}
one observes a clear plateau between $3\le t/a\le 5$ for the smaller value of $t_{\rm sep}$ 
which is in good agreement with the results shown in the left panel. 
In Figs.~\ref{fig:gem_pdep} the average values of the form factors are shown. 
As expected, in Fig.~\ref{fig:gem_pdep} the values for different $t_{\rm sep}$ agree 
within statistical errors, so we conclude that excited state contamination is small 
for $t_{\rm sep}\approx1.3-1.4$ fm source-sink separations used for the observables in this study.
A few percent precision on the form factors
for $G_e^p$, $G_m^p$ and $G_m^n$ is obtained, and less than 20\% precision for $G_e^n$. 
For $t_{\rm sep}=0.9$ fm even higher precision is seen despite having only a quarter of the statistics.
This indicates that $t_{\rm sep}=0.9$ fm allows good statistical precision 
while keeping control of excited state contamination. 

\subsection{EDM form factor}

\begin{figure}[tb]
\begin{center}
  \includegraphics[width=150mm]{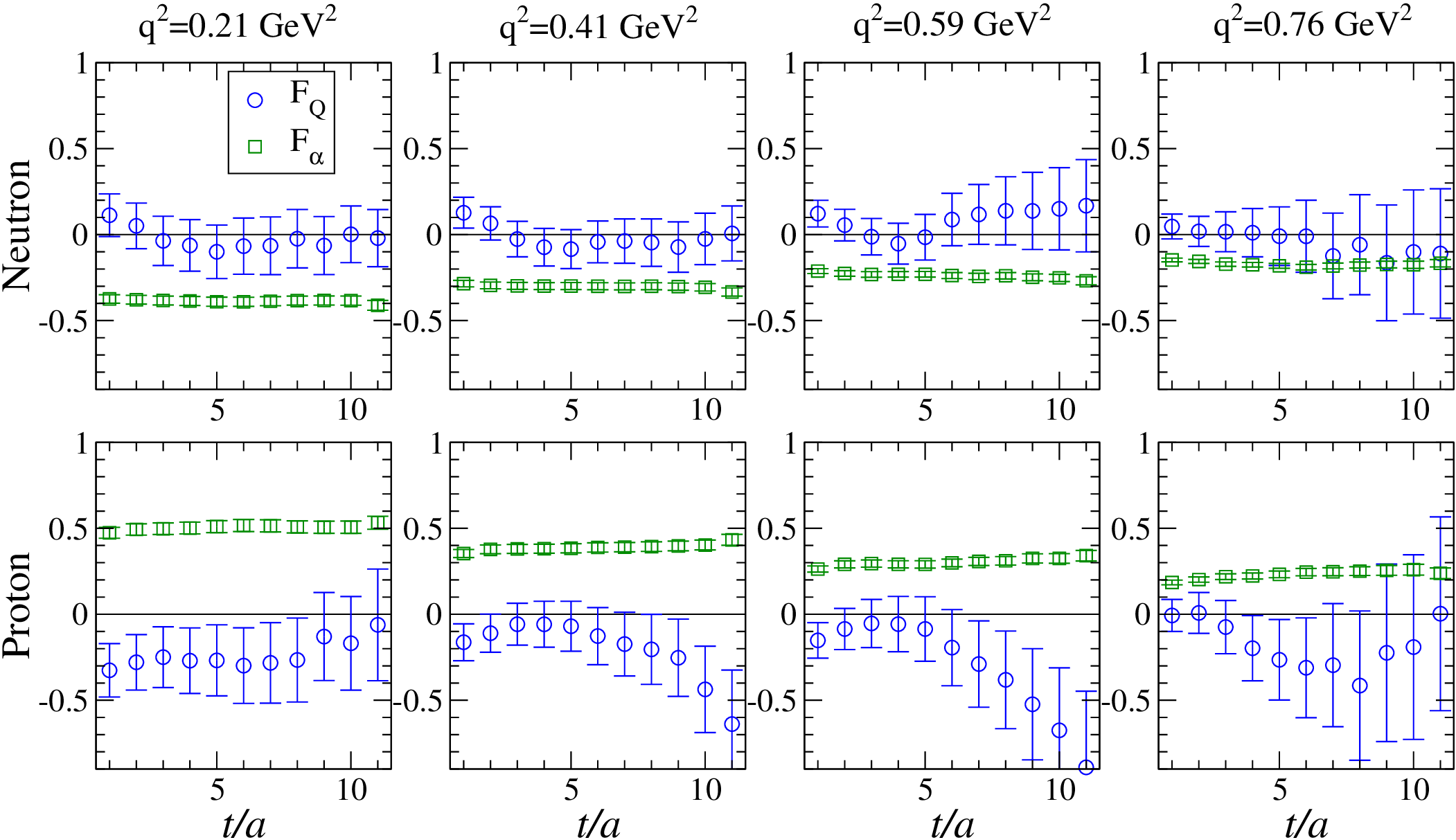}
  \caption{
  The operator time dependence for the components of the EDM form factor, $F_Q$ (total) and 
  the subtraction term $F_\alpha$.
  Momentum transfer increases from left to right. Iwasaki 24$^3$, 330 MeV pion ensemble.
  The three-point function is defined in (\ref{eq:rQ_t}).
  The source and sink operators are located in $t/a=0$ and 12. 
  }
  \label{fig:f3_tdep_sep}
\end{center}
\end{figure}

The EDM form factor is extracted from the CP-odd functions given in Eq.~(\ref{eq:rQ_t}) 
which contains $F_{3}$ and terms proportional to $\alpha$ to be subtracted.  
First we show decomposed $F_3$ into two pieces,
\begin{equation}
  F_3 = F_Q + F_\alpha,
  \label{eq:f3_sum}
\end{equation}
with
\begin{eqnarray}
  F_Q &=& \frac{m_N}{E_N+m_N}i\frac{2E_N}{p_z}
  {\rm tr}\Big[ P_{5z}^+\mathcal R^Q_{t}\Big],\\
  F_\alpha &=& \frac{m_N}{E_N+m_N}\alpha_N\Big(F_1 + \frac{3m_N+E_N}{2m_N}F_2\Big),
\end{eqnarray}
where $F_Q$ contains the total $\theta$-NLO three-point function, 
and $F_\alpha$ contains the subtraction terms.
From Figure \ref{fig:f3_tdep_sep}, one sees that $F_\alpha$ is relatively precise with a
statistical error of about 10\%,  while that of $F_Q$ is more than 50\%. 
This indicates that the ultimate signal-to-noise of $F_3$ depends mainly on $F_Q$. 
Again, the region $4\le t/a\le 8$ is used to obtain the EDM form factor. 

\begin{figure}[tb]
\begin{center}
  \includegraphics[width=150mm]{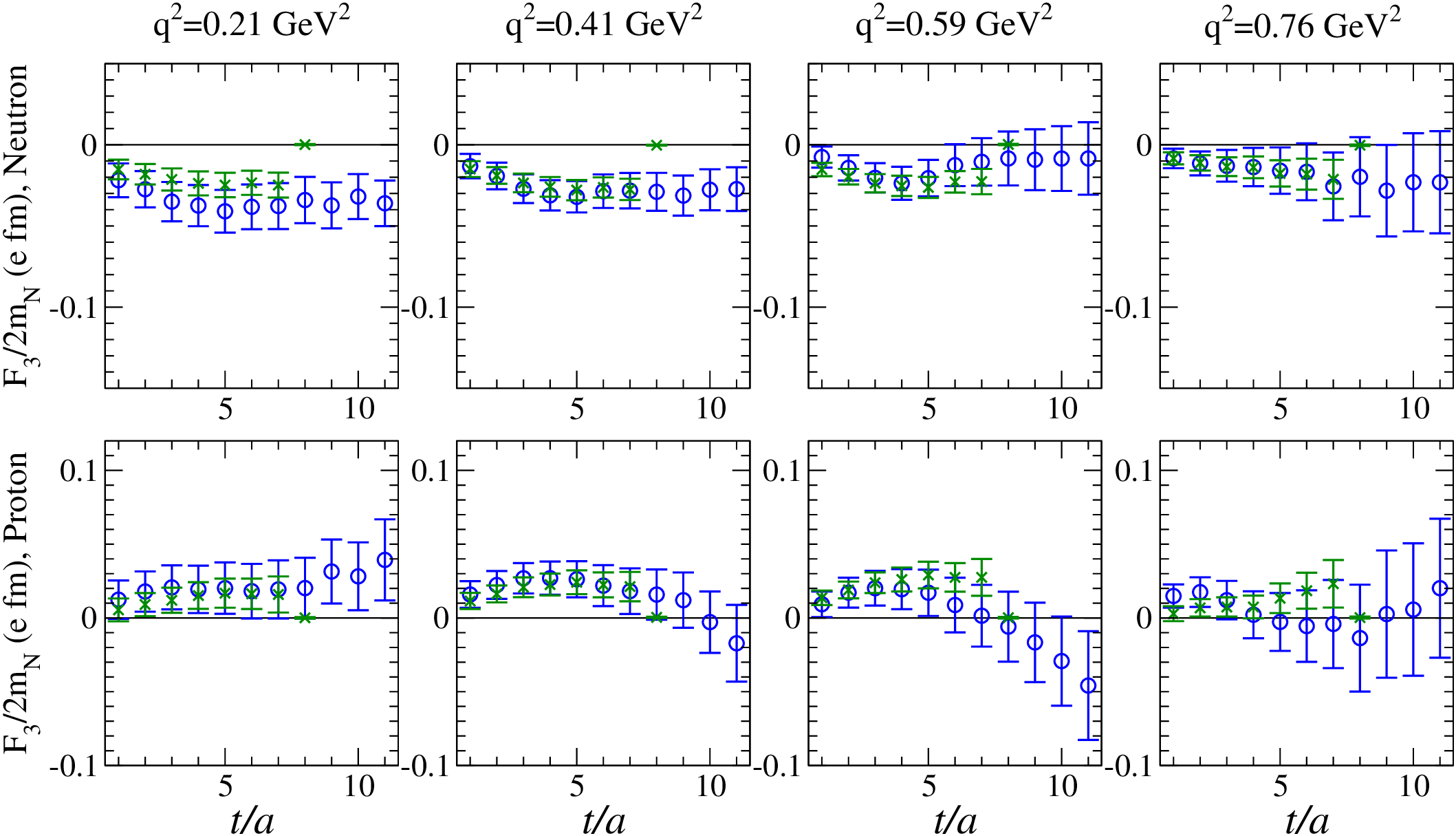}
  \caption{
  The EDM form factor for different source-sink separations. $t_{\rm sep}=1.32$ fm (circle) 
  and $t_{\rm sep}=0.9$ fm (cross), for neutron (top) and proton (bottom).
  Iwasaki 24$^3$, 330 MeV pion ensemble, at several momenta indicated in the above of each panel.
  We locate the source and sink operators in $t/a=0$ and 12 for $t_{\rm sep}=1.32$ fm,
  $t/a=0$ and 8 for $t_{\rm sep}=0.9$ fm.
  }
  \label{fig:f3_tdep_tsepdep}
\end{center}
\end{figure}

To investigate the presence of excited state contamination, 
we show the EDM form factor with $t_{\rm sep}=1.32$ fm and $t_{\rm sep}=0.9$ fm
in Fig.~\ref{fig:f3_tdep_tsepdep}.
The smaller separation result has an even better signal than $t_{\rm sep}=1.32$ fm, and 
their plateaus are consistent. 
Therefore one sees that the contamination of excited states is negligible in this range. 

\begin{figure}[tb]
\begin{center}
  \includegraphics[width=150mm]{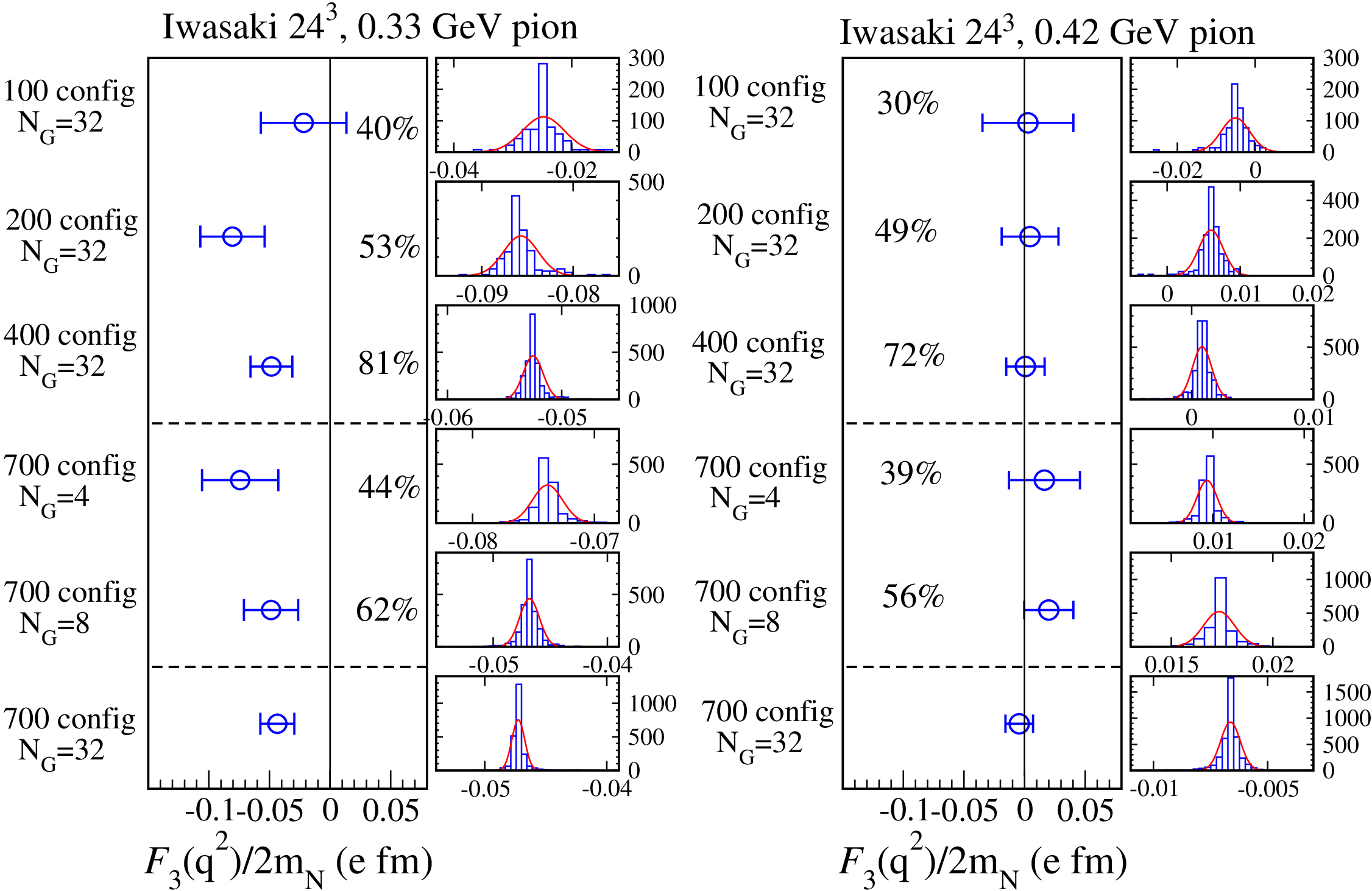}
  \caption{The neutron EDM form factor $F_3/2m_N$ in e$\cdot$fm unit,
    the lowest momentum, for various numbers of configurations and values of $N_G$. 
    The percentages denote the rates of reduction of statistical errors, defined as 
    the ratio of the statistical error 
    between full (bottom data) and reduced statistics cases. 
    The smaller panels show the distribution of jackknife estimates for each case. 
    The solid line denotes a Gaussian distribution function. 330 MeV pion (left)
    and 420 MeV pion (right) ensembles.
  }
  \label{fig:f3_statdep}
\end{center}
\end{figure}

In Fig.~\ref{fig:f3_statdep} we investigate statistical error scaling 
by examining subsets of our data and reduced $N_G$, the number of source locations of
$\mathcal O_G^{(\rm appx)}$ in the AMA procedure.
We find good agreement with the full results, and 
the statistical error roughly scales with the square root of the number of configurations.
Furthermore comparing the full statistics with reduced $N_{G}$, 
there is a similar reduction of the statistical errors,
{\it e.g.} the second line in Figure \ref{fig:f3_statdep} indicates 
the rate of $52$\% with one-quarter statistics (200 configurations) is close to 
the ideal rate, 50\%.
In the fourth line, 
the rate 44\% is slightly larger than the ideal rate $1/\sqrt{8}\simeq 35$\%.
It turns out that the gauge configurations we used do not show 
strong correlations between different trajectories, and also for AMA 
there is not a large correlation between different source locations.
Our choice of approximation and $N_G$
seem to perform well for the statistical error reduction of the EDM form factor for the Iwasaki $24^{3}$ ensembles, and
also we find that for the I-DSDR $32^{3}$ ensemble.

In Table \ref{tab:edmform_24c} and \ref{tab:edmform_ID}, we present the results of the EM 
and EDM form factors, extracted by fitting the plateaus to a constant value. 
The EDM form factors for the Iwasaki $24^{3}$ ensembles have roughly 25-30\% statistical errors, 
at best, and the errors grow to more than 100\% at worst, depending on the nucleon and momenta.
For the I-DSDR 32$^3$ lattice the EDM form factor is very noisy, and we do not observe a clear signal. 
This is likely due to the relatively poor sampling of the topological charge 
on this small ensemble of configurations since we do observe relatively 
small errors for the CP-even EM form factors.

In the next section we estimate the nucleon EDM's by extrapolating these results to zero momentum transfer. 

\begin{table}
\begin{center}
\caption{$F^n_3/2m_N$ (e$\cdot$ fm) on Iwasaki 24$^3$ ensemble.}
\label{tab:edmform_24c}
\begin{tabular}{c|cc|ccccc}
 \hline \hline
$m=0.005$ & P & & N \\
 \hline
$q^2$(GeV$^2$) & $t_{\rm sep}=1.32$ fm & $t_{\rm sep}=0.9$ fm & $t_{\rm sep}=1.32$ fm & $t_{\rm sep}=0.9$ fm\\
 0.210&   0.022(17)&  0.017( 9)& -0.040(13)&  -0.025( 7)\\
 0.405&   0.025(12)&  0.025( 7)& -0.031( 9)&  -0.027( 5)\\
 0.586&   0.013(15)&  0.028( 7)& -0.018(11)&  -0.026( 5)\\
 0.760&  -0.001(19)&  0.010( 7)& -0.018(14)&  -0.016( 6)\\
 \hline\hline
$m=0.01$ & P & & N \\
 \hline
$q^2$(GeV$^2$) & $t_{\rm sep}=1.32$ fm & $t_{\rm sep}=0.9$ fm & $t_{\rm sep}=1.32$ fm & $t_{\rm sep}=0.9$ fm\\
 0.212&   0.034(17)&   0.027(15)&  -0.005(11)&  -0.015(10)\\
 0.412&   0.023(13)&   0.021(11)&  -0.011( 8)&  -0.012( 7)\\
 0.604&  -0.006(15)&   0.014(10)&   0.003(10)&  -0.010( 7)\\
 0.782&   0.012(17)&   0.003( 9)&  -0.005(12)&  -0.002( 7)\\
 \hline\hline
\end{tabular}
\end{center}
\end{table}

\begin{table}
\begin{center}
\caption{$F^n_3/2m_N$ (e$\cdot$ fm) on I-DSDR, 32$^3$, 170 MeV pion ensemble.}
\label{tab:edmform_ID}
\begin{tabular}{c|c|cccccc}
 \hline \hline
  & P & N \\
 \hline
$q^2$(GeV$^2$) & $t_{\rm sep}=1.3$ fm & $t_{\rm sep}=1.3$ fm\\
 0.072&   0.033(80)&  -0.083(34)\\
 0.141&   0.057(50)&  -0.048(31)\\
 0.208&   0.027(69)&  -0.028(38)\\
 0.273&  -0.057(75)&  -0.067(50)\\
 \hline \hline
\end{tabular}
\end{center}
\end{table}

\subsection{Lattice results for the neutron and proton EDM}

To extrapolate to $q^2=0$ a simple linear function consistent with chiral perturbation theory is used,
\begin{equation}
  F_3(q^2)/2m_N = d_N + S^{\prime} q^2 + \mathcal O(q^4),
  \label{eq:f3fit}
\end{equation}
where $d_N$ represents the leading order, and $S^{\prime}$ the next-to-leading order 
in the $q^2$ dependence of the EDM form factor. 
$d_N$ is defined as the EDM. 
Furthermore, according to ChPT \cite{Kuckei:2005pg,Mereghetti:2010kp} at NLO, 
$S'$ in isoscalar (also isovector) is related to the low-energy constant
of CP violating pion-nucleon coupling, and this point is discussed later. 

\begin{figure}[tb]
\begin{center}
  \includegraphics[width=75mm]{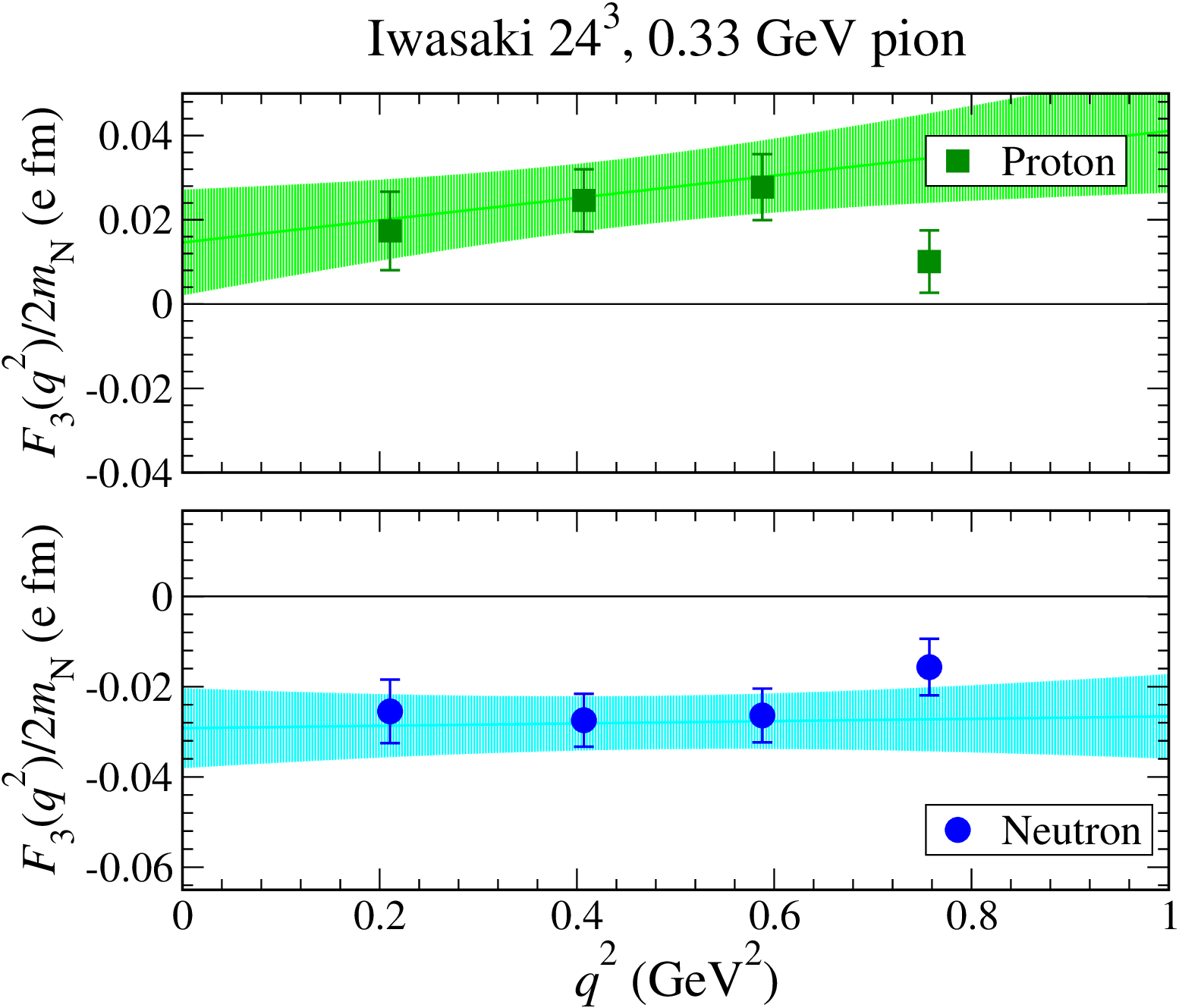}
  \vskip 5mm
  \includegraphics[width=75mm]{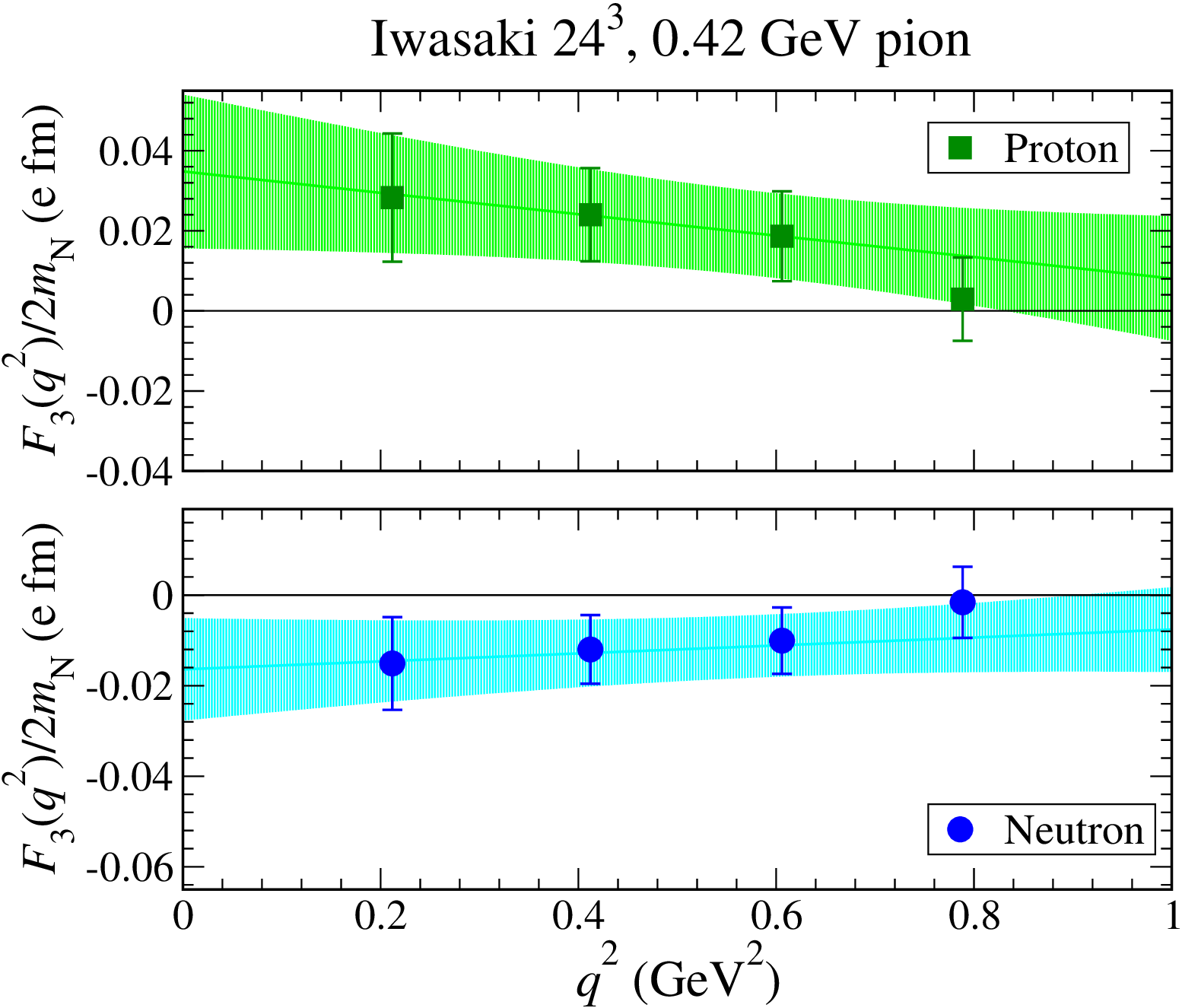}
  \vskip 5mm
  \includegraphics[width=75mm]{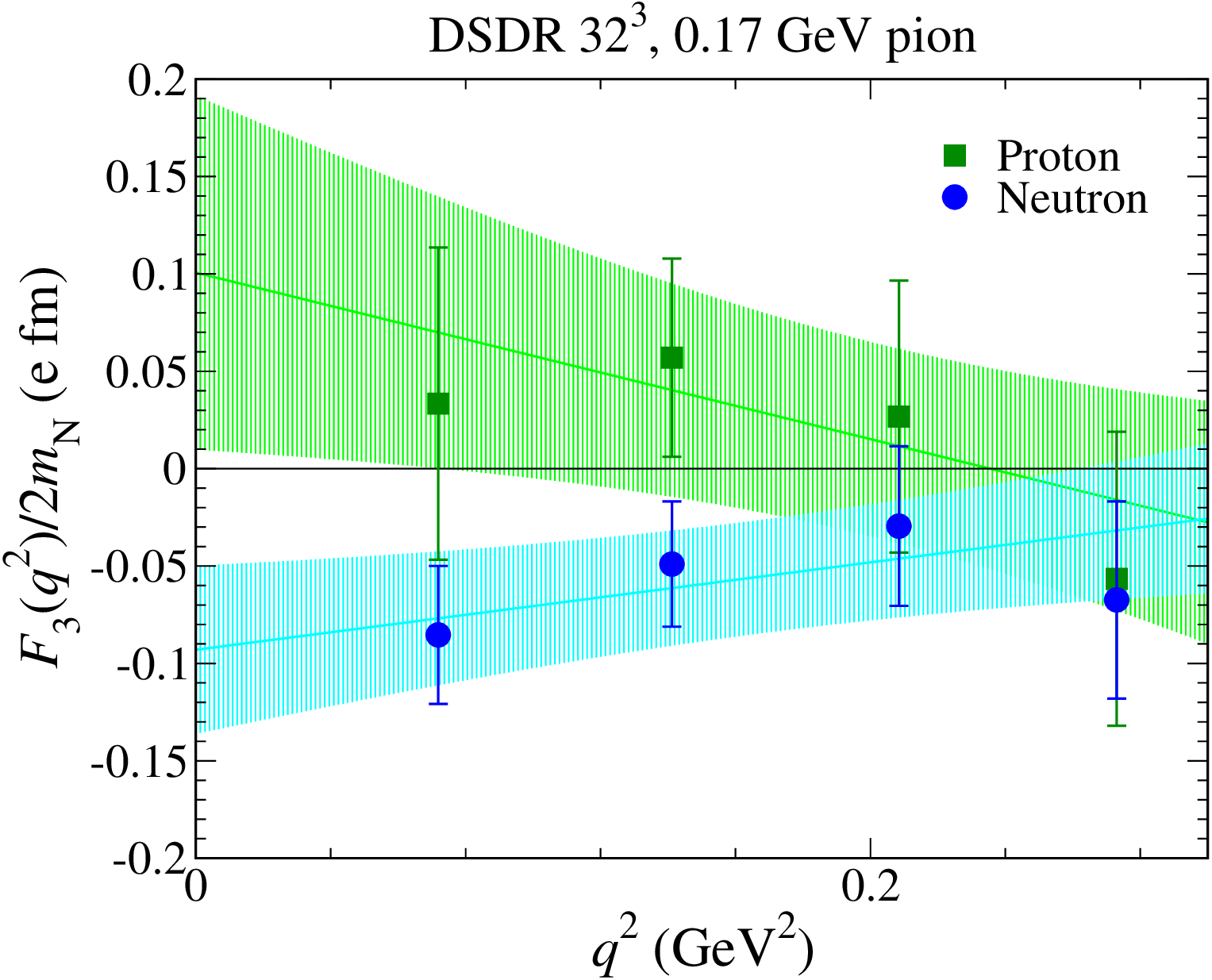}
  \caption{The EDM form factor for neutron (circle) and proton (square), 
  330 MeV (top) and 420 MeV (middle) pion, Iwasaki $24^3$ ensembles, 
  and 0.170 GeV pion (bottom), I-DSDR $32^3$ ensemble.
  In Iwasaki 24$^3$, $t_{\rm sep}=0.9$ fm is used. 
  The lines and bands denote the fitting function with statistical error. 
  }
  \label{fig:f3_q2dep}
\end{center}
\end{figure}

In Figs.~\ref{fig:f3_q2dep}, we show the $q^2$ dependence of the EDM form factors. 
$F_3(q^2)$ exhibits mild $q^2$ dependence within relatively large statistical errors.
Since we assume the linear function at low $q^2$ region for $F_3(q^2)$, 
fit ranges in low $q^2$, 0.20 GeV$^2< q^2<$ 0.6 GeV$^2$ in Iwasaki 24$^3$, 
and 0.07 GeV$^2< q^2<$ 0.273 GeV$^2$ in DSDR 32$^3$ are chosen. 
The central values and statistical errors for those fitting are given in Tab.~\ref{tab:edm}, 
and those lines and error bands are shown in Figure \ref{fig:f3_q2dep}.
One sees that using such fitting range, we have small $\chi^2$/dof, although 
the extrapolated EDM value has error of about 40--80\%, and also 
the slope of this function, which corresponds to $S'$, has almost 100\% statistical error. 
For the near physical pion mass ensemble the relative statistical error is still large:
the proton EDM is zero within one standard deviation and the neutron EDM is only non-zero
by a bit more than two.
Clearly more precision is needed.

\begin{figure}[tb]
\begin{center}
  \includegraphics[width=100mm]{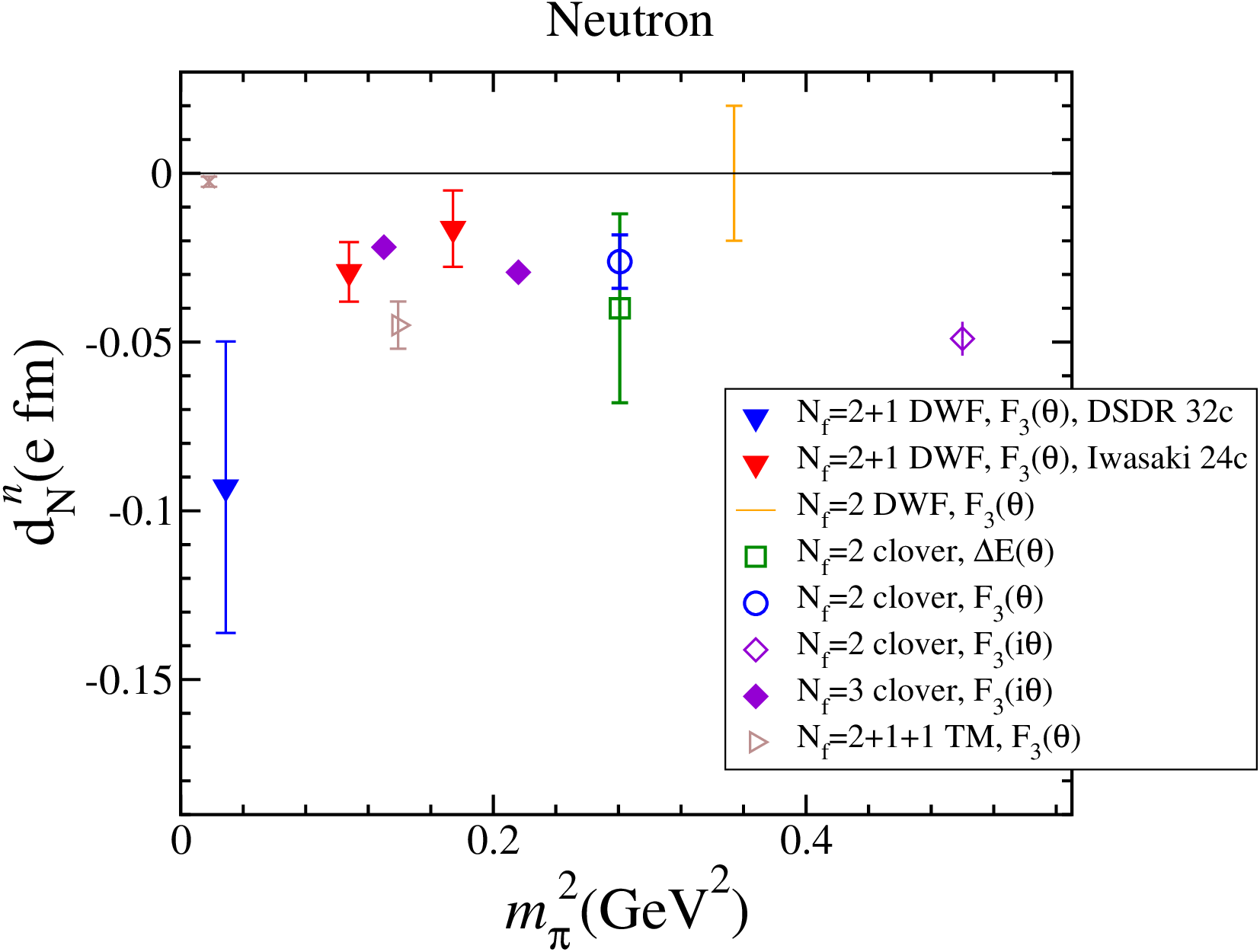}
  \vskip 5mm
  \includegraphics[width=100mm]{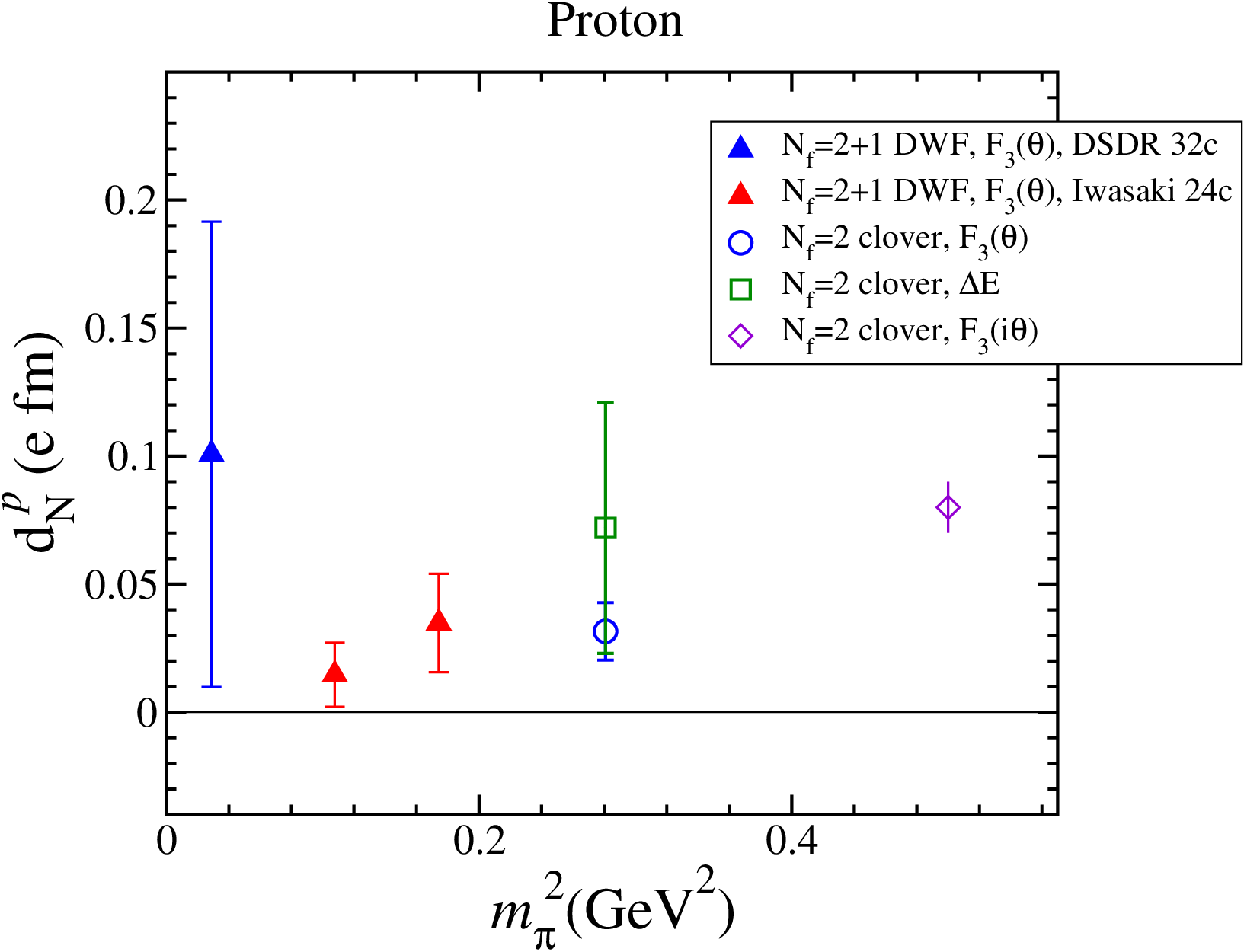}
  \caption{EDM summary plot for the neutron (top) and proton (bottom) 
    for 2 and 3 flavor QCD. Triangles denote results of the current study and
    include statistical and systematic errors, as described in the text. 
    Results for other methods are also shown: external electric field 
    ($\Delta E$) \cite{Shintani:ConfinementX2012},
    and imaginary $\theta$ ($F_3(i\theta)$)\cite{Aoki:2008gv,Guo:2015tla}.
    Previous results show statistical errors only.
    Right-triangle is result in $N_f=2+1+1$ TM fermion \cite{Alexandrou:2015spa}
    which is including systematic error. 
    The cross symbol in top panel denotes a range of values from model calculations of
    neutron EDM based on the baryon chiral perturbation theory
    \cite{Crewther:1979pi,Mereghetti:2010kp,Guo:2012vf}. 
  }
  \label{fig:dn}
\end{center}
\end{figure}

Figure \ref{fig:dn} displays our results for the EDM as a function of 
the pion mass squared, and for comparison we show older 
calculations with $N_f=2$ Wilson-clover and Domain-Wall fermions, and recent 
$N_f=3$ Wilson-clover fermions \cite{Guo:2015tla} and $N_f=2+1+1$ twisted-mass (TM) fermion 
\cite{Alexandrou:2015spa}.
One also sees that our results are comparable with
the recent imaginary-$\theta$ calculation\cite{Guo:2015tla} and ETMC collaboration \cite{Alexandrou:2015spa}. 
We note that DWF chiral symmetry forbids potentially large lattice artifacts
arising from mixing with chiral broken term 
associated with Wilson fermions \cite{Aoki:1990ix},
unlike the Wilson-clover simulations in \cite{Guo:2015tla}
(This corresponds to mixing term with topological charge and pseudoscalar mass term induced by lattice artifact.
Since in our case there is small residual mass which controls chiral symmetry breaking, then it is irrelevant in the current precision. 
However, if considering introducing the higher dimensional CP-violation operator, {\it e.g.} chromo-electric dipole moment, 
the mixing with lower-dimensional operator ($\theta$-term) should be taken into account, 
see \cite{Bhattacharya:2015rsa} for more details.).
Effective theories like chiral perturbation theory \cite{Crewther:1979pi,Mereghetti:2010kp,Guo:2012vf}
and several models in QCD sum rules \cite{Pospelov:1999mv,Hisano:2012sc}
have found $d_N^{p(n)}=(-)(1$--$4)\times 10^{-3}$ e$\cdot$fm
(the minus sign is for the neutron), about one order of magnitude smaller than the central value of lattice QCD results
computed at unphysically large pion mass.

\begin{table}
\begin{center}
  \caption{Result of EDM which is obtained by the extrapolation of $q^2$ to zero with linear ansatz
    using fitting range of 0.21 GeV$^2$ $\le q^2 \le 0.586$ GeV$^2$ for 24$^3$ m=0.005, 
    0.212 GeV$^2\le q^2\le 0.604$ GeV$^2$ for 24$^3$ m=0.01 and 
    0.072 GeV$^2\le q^2\le$ 0.273 GeV$^2$ for 32$^3$ DSDR m=0.001. 
    The value of $S'$ and its $\chi^2$/dof are also shown in this table. 
    Here those errors denote statistical one.}\label{tab:edm}
\begin{tabular}{cc|ccc|ccc}
 \hline \hline
Iwasaki 24$^3$ & & Proton & & & Neutron\\
 \hline 
$m_\pi$ (GeV) & $t_{\rm sep}$ (fm) 
    & $d_N^p$ (e$\cdot$fm) & $S^\prime_p$ (e$\cdot$fm$^3$) & $\chi^2$/dof
    & $d_N^n$ (e$\cdot$fm) & $S^\prime_n$ (e$\cdot$fm$^3$) & $\chi^2$/dof\\
\hline
0.33 & 1.32 &
0.030(25) & $-$11.0(21.2)$\times 10^{-4}$ & 0.7(1.7) & $-$0.053(18) & 24.3(14.6)$\times 10^{-4}$ & 0.2(9)\\
0.33 & 0.9 &
0.015(12) & 10.3(8.5)$\times 10^{-4}$ & 0.1(6) & $-$0.029(8) & 1.0(5.4)$\times 10^{-4}$ & 1.0(2.0)\\
0.42 & 1.32 &
0.064(27) & $-$45.2(21.8)$\times 10^{-4}$ & 1.3(2.3) & $-$0.021(15) & 11.7(12.9)$\times 10^{-4}$ & 1.8(2.7)\\
0.42 & 0.9 &
0.035(19) & $-$10.4(10.7)$\times 10^{-4}$ & 0.03(46) & $-$0.016(11) & 3.4(5.9)$\times 10^{-4}$ & 0.02(36)\\
 \hline \hline
 I-DSDR 32$^3$ & & Proton & & & Neutron\\
 \hline 
$m_\pi$ (GeV) & $t_{\rm sep}$ (fm) 
    & $d_N^p$ (e$\cdot$fm) & $S^\prime_p$ (e$\cdot$fm$^3$) & $\chi^2$/dof
    & $d_N^n$ (e$\cdot$fm) & $S^\prime_n$ (e$\cdot$fm$^3$) & $\chi^2$/dof\\
\hline
0.17 & 1.3 & 
0.101(90) & $-$166.4(147.1)$\times 10^{-4}$ & 0.4(7) & $-$0.093(43) & 87.4(74.0)$\times 10^{-4}$ & 0.5(9)\\
 \hline \hline
\end{tabular}
\end{center}
\end{table}

\section{Discussion}
\label{sec:discuss}

\begin{figure}[tb]
\begin{center}
  \includegraphics[width=110mm]{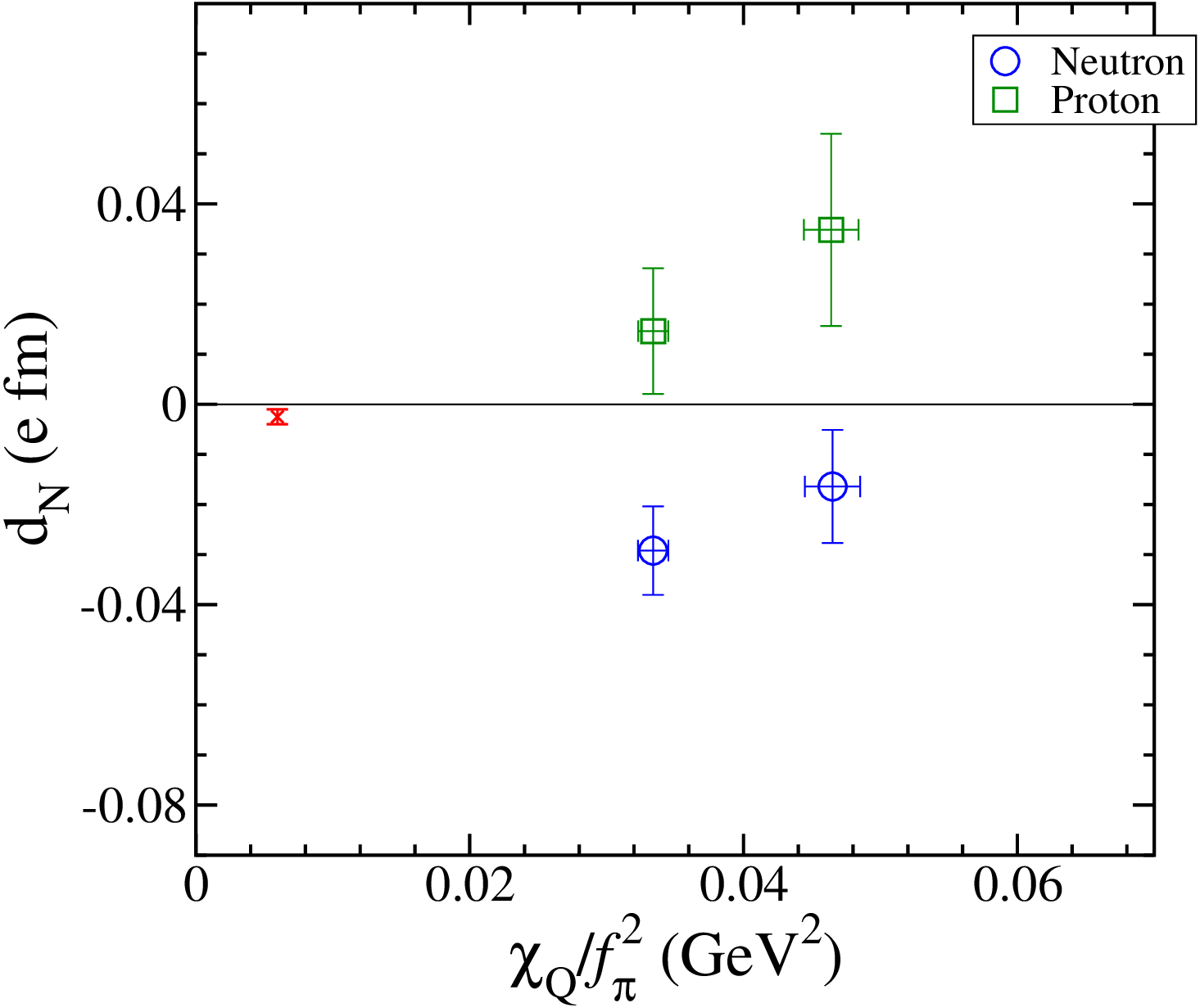}
  \caption{The relation between the nucleon EDM's and the topological charge susceptibility
    given in (\ref{eq:chpt_anom}) for the neutron (circle) and proton (square) in  
    Iwasaki $24^{3}$ ensembles.
    The cross symbol is value of neutron EDM from baryon chiral perturbation theory
    \cite{Crewther:1979pi,Mereghetti:2010kp,Guo:2012vf}.
  }
  \label{fig:edm_chidep}
\end{center}
\end{figure}

The neutron and proton EDM's induced by the $\theta$-term in the QCD action 
must vanish in the chiral limit since it can be moved entirely into a pseudoscalar mass term
by a chiral rotation because of the QCD axial anomaly
~\cite{Crewther:1979pi,Abada:1990bj,Aoki:1990zz,Cheng:1990pi,Abada:1991dv,Pich:1991fq,
  Borasoy:2000pq,O'Connell:2005un,Kuckei:2005pg,Ottnad:2009jw,Mereghetti:2010kp,Guo:2012vf}.
Such a mass term vanishes if any of the quarks in the theory are massless.
In chiral perturbation theory, the leading behavior~\cite{Crewther:1979pi} is
\begin{equation}
d_N \approx \frac{\bar g_{\pi NN} g_{\pi NN}}{m_N}\log \frac{m_{\pi}^{2}}{m_N^2}
\label{eq:chpt_anom}
\end{equation}
with CP-preserving and CP-violating $\pi$NN coupling,
$g_{\pi NN}$ and $\bar g_{\pi NN}$ respectively, 
whereas in the low energy nuclear effective theory \cite{Aoki:1990zz,Cheng:1990pi},
the EDM can also be described as 
\begin{equation}
d_N \approx\frac{2}{f_\pi^2}\chi_Q^2\mu_N \frac{\bar g_{\pi NN}}{2m_N}
\label{eq:chpt_anom2}
\end{equation}
where $\mu_{N}$ is the nucleon magnetic moment, $\chi_{Q}$ is the topological charge susceptibility,
which is represented in the leading order in chiral perturbation theory as 
$\chi_Q = m_\pi^2f_\pi^2(m_{\eta'}^2-m_\pi^2)/(N_fm_{\eta'}^2)$~\cite{Leutwyler:1992yt}
(here $f_\pi=0.092$ GeV). 
As given in Eq.~(\ref{eq:chpt_anom2}), topological charge distribution and its susceptibility is related to the EDM,
and thus it is interesting to see the relationship between $\chi_Q$ and EDM obtained in lattice QCD
for the consistency test with effective model. 
Figure \ref{fig:edm_chidep} shows such a relationship at our lattice point, 
and also displays the predicted bound from baryon ChPT at the physical point,
for which we use $m_\pi=0.135$ GeV and $m_{\eta'}=0.957$ GeV. 
One also sees that for the neutron EDM there is a slight tension between 
the lattice result and the ChPT estimate, however our simulation point 
is still far from the physical point.

Although the statistical uncertainty of our lattice results (Fig.~\ref{fig:dn})
is too large to discriminate the quark mass dependence
given in (\ref{eq:chpt_anom}) or (\ref{eq:chpt_anom2}), 
the sign of neutron and proton EDM's are opposite, and that sign is 
consistent with the nucleon magnetic moment as one can see in Fig.~\ref{fig:gem}.
Further, since the ratio of the proton and neutron EDM's is given from ratio of 
those magnetic moments as one can see in Eq.~(\ref{eq:chpt_anom2}), 
using quark model, its ratio is $(d_N^n/d_N^p)_{\rm quark} = -2/3$, assuming no SU(2) isospin breaking.
Our lattice calculation gives roughly 
$d_N^n/d_N^p\simeq -2$  and $d_N^n/d_N^p \simeq -0.5$ for the lighter
and heavier $24^{3}$ quark mass ensembles,
respectively, the same sign and order of magnitude as the quark model prediction.
Note that the analytic result of neutron EDM in NLO SU(2) \cite{Kuckei:2005pg} and SU(3) \cite{Ottnad:2009jw}
ChPT suggests that higher order corrections are about 40\%, 
and furthermore there is the additional uncertainty of the CPV $\pi NN$ coupling
\cite{Bsaisou:2014oka,Bsaisou:2014zwa,Mereghetti:2015rra}.


Nuclei or diamagnetic atoms ($e.g.$, $^{199}$Hg, $^{129}$Xe)
are important experimental avenues for detecting EDM's. 
To estimate their EDM's using an effective theory framework, non-perturbative evaluation of 
the low energy constants of the theory is essential. 
The low energy constants related to the quark mass and $q^2$ dependence of $F_3(q^2)$ and $S^\prime$,
for instance, can be obtained from lattice QCD.
The values of $S'$ in Tab.~\ref{tab:edm} (statistical errors only)
are similar order with the result of SU(3) ChPT at the leading-order, 
$S^\prime_{n}({\rm ChPT}) = -3.1\times 10^{-4}$ e$\cdot$fm$^3$ 
\cite{Kuckei:2005pg} (see also~\cite{Engel:2013lsa}).
\red{
  Furthermore, according to the argument of NLO BChPT (for details, see \cite{Mereghetti:2015rra}),
  $S'$ for the isoscalar and isovector EDMs, 
  is approximately 
  \begin{equation}
    S'_{\rm isoscalar} \simeq 0, \quad
    S'_{\rm isovector} \simeq \frac{g_A\bar g_\pi^{(0)}}{24\pi^2f_\pi m_\pi^2}
    \Big[ 1 - \frac{5\pi}{4}\frac{m_\pi}{m_N}\Big],
  \end{equation}
  so $\bar g_\pi^{(0)}$, the CPV $NN\pi$ coupling, is leading in $S'_{\rm isovector}$.
  Although the precision shown in Tab.~\ref{tab:edm} is not enough to address this comparison,
  our results provide a rough bound, $|\bar g_\pi^{(0)}|\sim O(10^{-1})$.
  The phenomenological value is also estimated as $\bar g_\pi^{(0)}\sim 0.04$ \cite{Engel:2013lsa}.
}

\begin{figure}[tb]
\begin{center}
  \includegraphics[width=110mm]{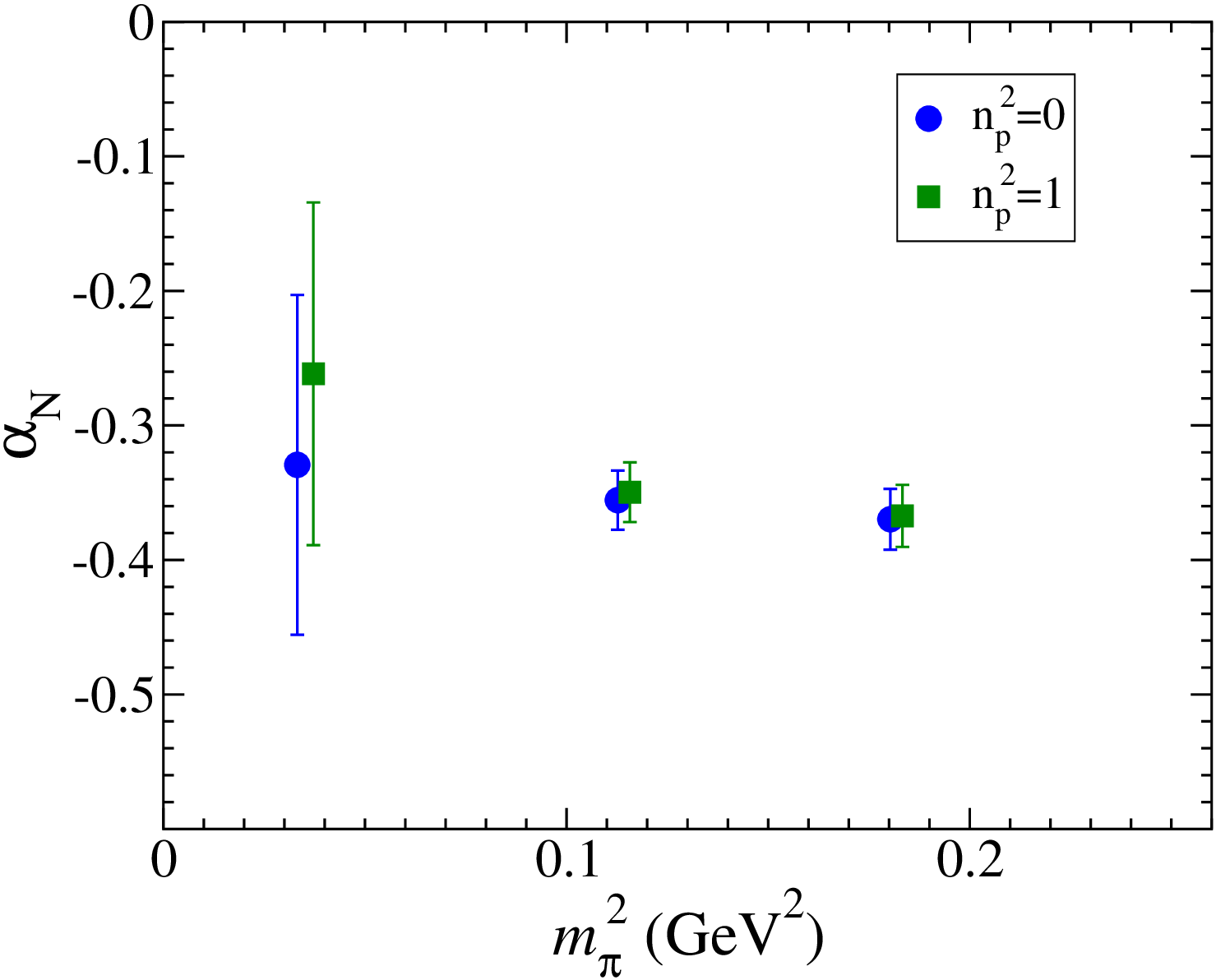}
  \caption{The dependence of pion mass squared for $\alpha_N$ obtained by CP-odd nucleon two-point function 
    using the different momenta.
  }
  \label{fig:alphaN_mdep}
\end{center}
\end{figure}

Finally we consider the chiral behavior of the CP-odd mixing angle $\alpha_{N}$.
It depends on the (sea) quark mass but is independent of momentum.
Since $\alpha_{N}(\theta)\propto\theta$, it is expected to vanish in the chiral limit.
However, as seen in Fig.~\ref{fig:alphaN_mdep}, we observe no significant 
mass dependence for $\alpha_N$ among all of the ensembles in our study.  
This may simply reflect that the simulations are far from the chiral limit for EDM's.
We also note that the statistical errors are large, especially for the 170 MeV pion ensemble,
and there the topological charge distribution is suspect since we have only used 39 configurations.

\section{An exploratory reweighting with topological charge density}
\label{sec:idea}

\begin{figure}[h]
\begin{center}
  \includegraphics[width=100mm]{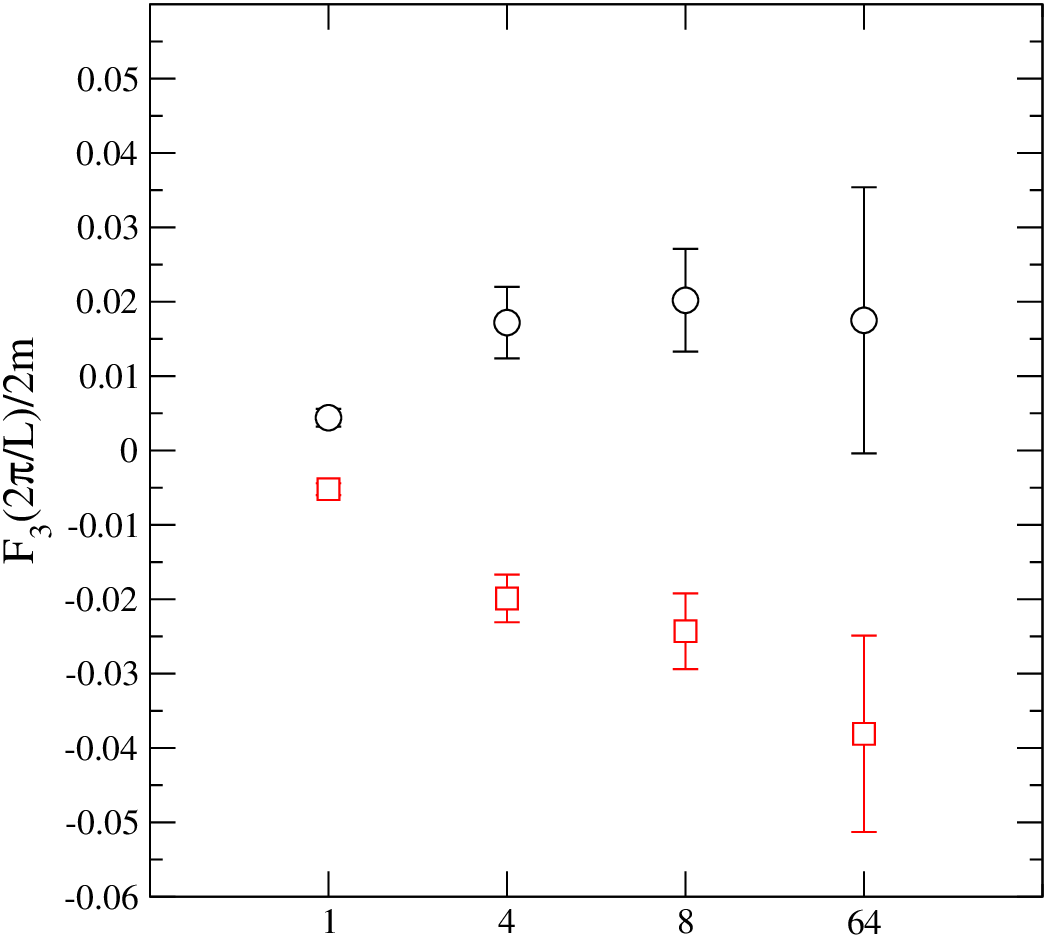}
  \vskip 5mm
  \includegraphics[width=100mm]{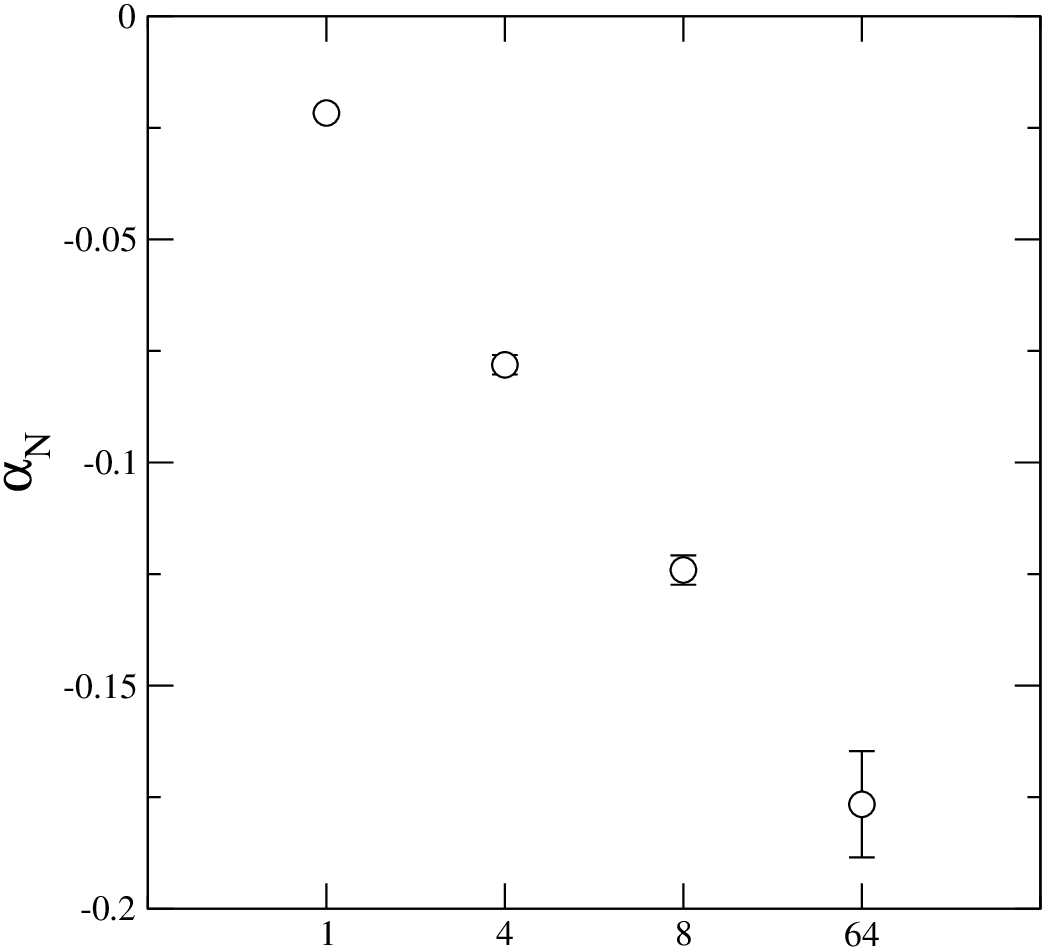}
  \caption{(Top) The nucleon EDM form factors from local time slice reweighting, 
    as described in the text, for the lowest non-trivial momentum. Proton (squares) and neutron (circles). 
    The point on the right corresponds to reweighting with the topological charge $Q$. 
    $24^{3}$, 330 MeV pion ensemble.
    (Bottom) CP-odd mixing angle from local time slice reweighting, as described in the text, 
    on the same ensemble.
  }
  \label{fig:local reweight}
\end{center}
\end{figure}

Large statistical noise of CP-odd correlation function is possibly due to reweighting with the global topological charge
since for many, perhaps most, of the current insertions, there is no overlap with a CP-odd vacuum fluctuation, 
so reweighting just adds noise to the expectation value. 
Unfortunately for this study, we have averaged over space on each time slice, 
so we can not examine these local correlations directly. 
But we can reweight the correlation function with the charge density summed over a time slice, or 
several successive time slices. 
To investigate the above, we sum the topological charge density over a range of time slices, 
1, 4, 8 (which is corresponding to temporal location of sink operator) and 64 (which is the maximum size of temporal extension), 
symmetrically straddling the EM current insertion on a given time slice. 
A plot of the nucleon EDM for such a reweighting is shown in Fig.~\ref{fig:local reweight}, 
and the corresponding mixing angle.

One observes a dramatic decrease in the noise as the number of time slices that are summed for the topological charge density decreases. Interestingly, the values appear to reach a plateau between 9 and 17 time slices. In the future, we plan to investigate spatially local reweighting. One needs to address issues of renormalization as well.
  

\section{Summary}
\label{sec:summary}

This paper presents a lattice calculation of the nucleon electric dipole moment obtained from the study of the CP-odd form factors of the nucleon in 2+1 flavor QCD with unphysically heavy up and down quarks  (the pion mass in this study ranges from 420 down to 170 MeV).  
The QCD $\theta$-term is included to the lowest order by reweighting correlation functions with the topological charge. 
We employ the domain wall fermion discretization of the lattice Dirac operator which allows us to control lattice artifacts due to 
chiral symmetry breaking which may otherwise lead to significant systematic errors in the chiral regime. 
We applied the all-mode-averaging (AMA) procedure \cite{Blum:2012uh,Blum:2012my}
to significantly boost the statistical precision of the correlation functions which resulted in 
statistically significant values of the neutron and proton EDM's for the two heavier quark ensembles in our study, 
and a less significant signal for the lightest, 170 MeV pion ensemble. 
We have examined the pion mass dependence of the EDM's, which is obtained by linear extrapolation of 
low momentum transfer to zero momentum transfer with two different time-slice separation 
of source and sink operators.
In this analysis, the effect of excited state contamination is small compared to the statistical error.

In addition, we have investigated the relationship between the local topological charge on each time slice of the lattice
and the CP-odd correlation function. This idea may lead to a significant noise reduction in future calculations by reweighting correlation functions
with the local topological charge density.  
We show promising numerical evidence that the large noise associated with global topological charge fluctuations can be reduced.

In this paper, we have concentrated on a high statistics analysis using unphysical masses, 
$m_\pi=$0.17 GeV -- 0.42 GeV, and provide lattice QCD results for the nucleon EDMs and form factors 
with statistical errors only. 
Future calculations will address systematic errors, including finite size effects (FSE), 
poor topological charge sampling, the $q^2=0$ extrapolation, and lattice spacing artifacts. 
Baryon chiral perturbation theory (BChPT) in finite volume, to the next-to-leading order
\cite{O'Connell:2005un,Guo:2012vf,Akan:2014yha},
suggests the magnitude of FSE for our lattice sizes and pion masses are roughly 10\%, or less. 
However additional effects are possible, for instance, at higher order in BChPT. 
We note several domain-wall fermion gauge ensembles 
with different lattice cutoffs, volumes and pion masses below 0.2 GeV are available
~\cite{Arthur:2012opa,Blum:2014tka} to estimate these systematics. 
Recent developments in numerical algorithms like AMA make it possible to carry out these calculations with current 
computational resources, and those studies are under way. 

\begin{acknowledgments}
  We thank members of RIKEN-BNL-Columbia (RBC) and UKQCD collaboration for sharing USQCD resources for part of our calculation.
  ES thanks F.-K.~Guo and U.-G.~Meissner, E.~Mereghetti, J.~de~Vries, U.~van~Kolck and
  M.~J.~Ramsey-Musolf for useful discussions on chiral perturbation theory, 
  and also G.~Schierholz, A.~Shindler for discussion and comments. 
  Numerical calculations were performed using the RICC at RIKEN and the Ds cluster at FNAL.
  This work was supported by the Japanese Ministry of Education Grant-in-Aid,
  Nos. 22540301 (TI), 23105714 (ES), 23105715 (TI) and U.S. DOE grants DE-AC02-98CH10886 (TI and AS)
  and DE-FG02-13ER41989 (TB).
  We are grateful to BNL, the RIKEN BNL Research Center, RIKEN Advanced Center for Computing and Communication, 
  and USQCD for providing resources necessary for completion of this work.
  For their support, we also thank the INT and organizers of Program INT-15-3
  ``Intersections of BSM Phenomenology and QCD for New Physics Searches'', September 14 - October 23, 2015.
\end{acknowledgments}

\bibliography{ref}
\end{document}